\documentstyle[11pt]{article}
\begin{document}
\begin{flushright}
DO-TH-95/17 \\
(hep-ph/9509350)
\end{flushright}
\centerline{\Large \bf Top quark effects on the scalar sector}
\centerline{\Large \bf of the minimal Standard Model}
\vspace{1.2cm}
\begin{center}
{{\bf G.~Cveti\v c} \\
Inst.~f\"ur Physik, Universit\"at Dortmund, 44221 Dortmund, Germany}\\
\end{center}

\vspace{1.2cm}
\centerline{\bf Abstract}
We study effects of the heavy top quark ($m_t \approx 180 \mbox{GeV}$)
on the scalar sector of the minimal Standard Model. We construct the
effective potential for the scalar doublet, by first taking into
account the leading contributions of the top quark loops. Minimizing
this potential gives us a condition analogous to the leading-$N_c$ gap
equation of the standard $\langle {\bar t} t \rangle$-condensation
model (Top-mode Standard model). This essentially non-perturbative
condition leads to a low ultraviolet cut-off $\Lambda = {\cal {O}}
(1 \mbox{ TeV})$ in the case when the bare mass $\mu$ of the scalar
doublet in the tree-level potential satisfies $\mu^2 = -M^2_0 \leq 0$
and the scalar doublet there is self-interacting ($\lambda > 0$). We
demand that the scalar self-interaction behave perturbatively -- in
the sense that its 1-loop contributions influence the effective
potential distinctly less than those of the Yukawa coupling of the
heavy top quark. When we subsequently include the 1-loop contributions
of the scalar and the gauge bosonic sectors in a perturbative manner,
the results change numerically, but the cut-off $\Lambda$ remains
${\cal {O}}(1 \mbox{ TeV})$. The resulting Higgs mass $M_H$ is then in
the range 150-250 GeV. Furthermore, the results of the paper survive
even in the case when the square of the bare mass $\mu^2$ is positive,
as long as $\mu^2 \leq {\cal {O}}(\lambda v^2)$, where $\lambda/4!$ is
the usual bare coupling parameter of the quartic self-interaction of
the scalars and $v$ is the vacuum expectation value ($v=246.2 \mbox{
GeV}$,  $M^2_H \approx \lambda v^2/3$). \\
PACS number(s): 11.30.Qc, 11.15.Ex, 12.15.Lk, 14.80.Bn

\vspace{1.cm}

\section{Introduction}

In the minimal Standard Model (MSM), the scalar sector is probably the 
most misterious one. Experiments haven't yet shown any direct evidence
of the Higgs. The indirect evidence is very hard to pinpoint, since
the results of the measurements up to date either do not depend or
depend very mildly on the scalar structure of the model and on the Higgs 
mass. Therefore, the mass of the Higgs of the MSM is still largely 
unknown ($65 \mbox{ GeV} < M_H \stackrel{<}{\sim} 0.8 \mbox{TeV}$)~\cite{
sopczak}. The main reason for having the scalar sector in the Standard 
Model is to have a viable mechanism to induce an electroweak spontaneous
symmetry breaking (SSB) leading to the phenomenologically required masses
of the electroweak gauge bosons $W$ and $Z$. While SSB in this scenario
is directly related to parameters of the scalar sector, these can be
dramatically influenced by effects of other sectors on the scalar one.

The experimental evidence of a heavy top quark ($m_t \approx 180
\mbox{ GeV}$)~\cite{CDF} suggests that the resulting strong Yukawa
coupling parameter $g_t \approx 1$ may drastically
influence the scalar sector, through quantum loops. These effects
on the parameters of the effective potential of the scalar doublet
$V_{\mbox{\footnotesize eff}}(\Phi) $ may have a non-perturbative nature. 
For example, they may be responsible for inducing SSB.
To treat such non-perturbative effects with 
diagrammatic (loop) approach is legitimate, because these are
effects of one (quark) sector of the model on another (scalar) sector.
On the other hand, the diagrammatic treatment of the quantum effects 
of the scalar self-interaction on the scalar sector itself has predictive
power only if these effects are reasonably weak -- of perturbative
nature. 

In section II, we investigate the leading effects of the heavy 
top quark sector on the scalar sector, by calculating the contributions
of the top quark loop to the effective potential 
$V_{\mbox{\footnotesize eff}}(\Phi)$. Minimizing this potential, we
end up with a relation connecting the bare parameters of the scalar 
sector~\footnote{
i.e, the bare parameters {\it before} the inclusion of the top quark
effects.} 
to the ultraviolet cut-off, the vacuum expectation value and $m_t$.
This relation is analogous to the leading-$N_c$ gap equation in
a simple model of the $\langle {\bar t} t \rangle$-condensation
(Top-mode Standard Model - TSM~\cite{bhl}). From this relation we infer
that the cut-off $\Lambda$ of the theory should have a stringent upper
bound, if the square of the bare mass $\mu^2$ of the scalar doublet
is non-positive~\footnote{
Note that for negative $\mu^2$ ($\mu^2 \not= 0$) we have SSB already
at the tree level.}
or at least smaller than ${\cal {O}} (\lambda v^2)$, where
$\lambda/4!$ is the bare coupling parameter of the quartic scalar
self-interaction ($\lambda \geq 0$), and $v=$246 GeV. Under these
conditions, we obtain the upper bound $\Lambda^{\mbox{\scriptsize u.b.}}$
and the Higgs mass $M_H$
as a function of the parameter $\lambda$. For $\lambda \stackrel{<}
{\approx} 3$, we obtain $\Lambda^{\mbox{\scriptsize u.b.}} = {\cal {O}}
(1 \mbox{ TeV})$ and $M_H$ in the region of 150-250 GeV.
If, however, $\mu^2$ is positive
{\it and} surpasses $\lambda v^2/6$, the cut-off $\Lambda$ of the 
theory becomes less restricted and values 
$\Lambda \geq {\cal{O}}(10 \mbox{ TeV})$ become possible.
In such a case, since $\mu^2 \geq 0$, we do not have SSB at the tree level.

In Section III, we include in a perturbative manner the 1-loop
effects of the scalar self-interactions, assuming them not to be 
too strong. By this we mean that they are distinctly smaller
than the heavy top quark effects calculated in Section II.
This in turn implies that the value of the bare coupling parameter
$\lambda$ cannot be larger than about 3.
We also include the 1-loop effects of the gauge bosons. The latter
effects are substantially smaller than the contributions of the heavy
quark sector, and are hence also calculated in a perturbative manner. 
The modified numerical results are presented in Table 1. It turns out 
that the qualitative features of Section II survive, i.e., $\Lambda
= {\cal {O}}(1 \mbox{ TeV})$ and $M_H = 150-250 \mbox{ GeV}$.

For the results of the present paper, it was important to treat the
cut-off $\Lambda$ of the model (MSM) as a finite and physical quantity,
i.e., $\Lambda$ is roughly the energy where the MSM is replaced by
some new, as yet unknown, physics. For simplicity, we were using simple
covariant spherical cut-off for the Euclidean 4-momenta of the loops
(after the Wick rotation). The presence of $\Lambda^2$ and $\ln
(\Lambda/m_t)$-terms in the effective potential was crucial in
our analysis. We neglected terms of ${\cal {O}}(\Lambda^0)$.
The possible errors resulting from this and other
approximations were estimated toward the end of the paper.
We emphasize that the terms ${\cal {O}}(\Lambda^0)$ in our integrals
depend on the regularization (cut-off) procedure chosen. 

We stress that the present work was largely motivated by the
work of Fatelo {\it et al.}~\cite{fghw} and the ideas contained
therein. They discussed the case
of the zero bare coupling $\lambda$. The present work can be
regarded as an extension of their work to the case of nonzero
values of the bare coupling $\lambda$.

The basic result of the present paper is the following:
in a substantial part of the parameter space of the bare scalar 
couplings, the interplay of the heavy top and the scalar sector leads 
to the conclusion that the MSM is replaced by some new physics
at a relatively low scale ${\cal {O}}(\mbox{1 TeV})$. In other
parts of that parameter space, much larger cut-offs are
still possible. The results are valid and predictable as long as
the scalar sector is not too strongly self-interacting, i.e., as
long as the Higgs mass is in the region 150-250 GeV.

\section{Top quark (non-perturbative) contributions to
the effective potential}

We start with the following Lagrangian, containing only the
sectors of our primary interest -- the scalar and the quark
sector, assuming that the only non-zero Yukawa coupling is $g_t$:
\begin{eqnarray}
{\cal {L}}^{(\Lambda)} & = & i {\bar {t^a}}{\partial \llap /} t_a
+ i {\bar {b^a}}{\partial \llap /} b_a
+  \partial_{\mu}\Phi \partial^{\mu}\Phi^{\dagger}
- V^{(0)} \left( 2 \Phi\Phi^{\dagger}; \Lambda \right)
\nonumber\\
& & - \frac{g_t\left( \Lambda \right)}{\sqrt{2}} {\bar {t^a}} \left(
\varphi - i \gamma_5 G^{(0)} \right) t_a
+ \frac{g_t\left( \Lambda \right)}{2} \left[ G^{(+)}{\bar {t^a}}
\left( 1 - \gamma_5 \right) b_a +
G^{(-)}{\bar {b^a}} \left( 1 + \gamma_5 \right) t_a \right] \ ,
\label{Lagr}
\end{eqnarray}
where we took $g_b(\Lambda) \approx 0$, and $\Phi$ is the
scalar $SU(2)_L$-doublet of the MSM
\begin{equation}
\Phi = \frac{1}{\sqrt{2}} { \sqrt{2} G^{(+)} \choose \varphi + i G^{(0)} } \ .
\label{Phi}
\end{equation}
Here, $\varphi$, $G^{(0)}$ and $G^{(\pm)}$ are the Higgs field (before the
symmetry breaking) and the neutral and the charged Goldstone fields,
respectively. In (\ref{Lagr}), $\Lambda$ is the
effective cut-off of the theory, $a$ is the color index for the
top quark, and $V^{(0)}$ denotes the tree level potential
\begin{equation}
V^{(0)}\left( \varphi^2; \Lambda \right)
= -\frac{1}{2} M^2(\Lambda) \varphi^2
+ \frac{\lambda\left( \Lambda \right) }{4!}
  \varphi^4 \ .
\label{Vtree}
\end{equation}
Here, and from now on, we denote in the effective potentials explicitly
only the dependence on the square $\varphi^2$ of the unbroken Higgs field.
The general $SU(2)_L$-invariant expressions are obtained always by
simply substituting $\varphi^2 \mapsto 2 \Phi \Phi^{\dagger}$, thus
explicitly including the Goldstone degrees of freedom. 
In (\ref{Vtree}), $M^2(\Lambda) = - \mu^2(\Lambda)$, where $\mu(\Lambda)$ 
is formally the bare mass of the scalars before the electrowek
symmetry breaking; $\lambda(\Lambda)$ is the
non-negative bare parameter of the quartic self-coupling
term.  For simplicity, we always omit the superscript $\Lambda$ in the bare
fields: $\varphi^{(\Lambda)} \mapsto \varphi$. 
In this Section, where we only consider the leading
influence of the heavy top quark on the scalar sector,
the Goldstone degrees of freedom will not play any direct role.

It is well known that in the case of ``imaginary'' masses $\mu$, i.e.,
$M^2(\Lambda) > 0$, we have spontaneous symmetry breaking (SSB)
already at the tree level: $\langle \varphi \rangle =
\pm \sqrt{ 6 M^2(\Lambda)/\lambda(\Lambda) }$ and $M_H^2(\Lambda) = 
\lambda(\Lambda) \langle \varphi \rangle^2/3$, where $\langle \varphi
\rangle$ means the vacuum expectation value (VEV), and $M_H$ is the
mass of the Higgs $H=\varphi - \langle \varphi \rangle$.

The heavy top quark contributes at 1-loop level appreciably
to the effective potential of the Higgs, because of the strong Yukawa 
coupling $g_t(\Lambda) \sim 1$. These contributions can be obtained,
for example, by simply calculating the truncated Green functions
${\tilde \Gamma}_{\varphi}^{(2n)} \left( p_1, \ldots, p_{2n} \right)$ 
at zero external momenta, corresponding to the diagrams of Figs.~1a-c
\begin{equation}
V_{\mbox{\footnotesize eff}} \left( \varphi^2; \Lambda \right) = 
V^{(0)}\left(\varphi^2; \Lambda \right) +
i \sum_{n=1}^{\infty} \frac{1}{\left( 2 n \right) !}
{\tilde \Gamma}_{\varphi}^{(2n)} \left( p_1, \ldots, p_{2n} \right)
{\Big |}_{ ( \lbrace p_k \rbrace = 0 ) } {\varphi}^{2n} \ .
 \label{V1lt}
\end{equation}
This was calculated, for example, in ref.~\cite{gceap}, for the case
of a scalar field
which is initially unbroken and non-dynamical (auxiliary), within a 
$\langle \bar t t \rangle$-condensation framework 
(Top-mode Standard Model -- TSM).
It is straightforward to see that the question whether the scalar $\varphi$
is dynamical or non-dynamical at the outset, does not play any role
for the truncated Green functions corresponding to Figs.~1a-c. Furthermore,
we have explicitly checked that the 1-loop contribution of the top 
quark to $V_{\mbox{\footnotesize eff}}$ remains the same
even when the field $\varphi$ is broken already at the tree 
level~\footnote{
This conclusion could be arrived at also by invoking 
$SU(2)_L$-symmetry arguments.}, i.e., when $M^2(\Lambda) > 0$.
Therefore, we can simply copy the result of ref.~\cite{gceap}, replacing
there the formal Yukawa coupling parameter $g = M_0 \sqrt{G}$ by the 
actual bare Yukawa coupling parameter $g_t(\Lambda)$
\begin{eqnarray}
V_{\mbox {\footnotesize eff}}\left( \varphi^2; \Lambda \right) & = &
V^{(0)}\left( \varphi^2; \Lambda \right) +
V^{(1\ell t)}\left( \varphi^2; \Lambda \right) \nonumber\\
&  = & V^{(0)}\left( \varphi^2; \Lambda \right) -
\frac{N_c}{8 \pi^2} \int_0^{\Lambda^2} d{\bar k}^2 {\bar k}^2
\ln \left[ 1 + \frac{ g^2_t(\Lambda) \varphi^2}{2 {\bar k}^2} \right] \ .
\label{V1l}
\end{eqnarray}
We see that the 1-loop top quark (superscript: $1\ell t$)
contribution is proportional to the 
number of colors $N_c=3$, since each color contributes independently 
to the loops of Figs.~1a-c. For the integral (\ref{V1l}), Wick rotation
has been performed and the integral
is written in the Euclidean metric, ${\bar k}$ is the Euclidean loop momentum
of the top quark. For simplicity, we used covariant spherical cut-off.

The minimum of $V_{\mbox {\footnotesize eff}}$ is achieved at
the value of $\varphi$ that is the vacuum expectation value
(VEV) $\langle \varphi \rangle_{1\ell t}$,
where the subscript denotes that this is an approximation with
only the leading (1-loop) heavy top quark quantum effects taken into
account. We will call the relation resulting from this minimum 
the ``gap'' equation~\footnote{
We choose this term in analogy with the terminology of the
$\langle \bar t t \rangle$-condensation mechanism.}
\begin{displaymath}
\frac{\partial V_{\mbox{\footnotesize eff}}^{(0+1\ell t)} 
\left( \varphi^2; \Lambda \right)}{\partial \varphi^2} \Bigg|_{\varphi
= \langle \varphi \rangle_{1\ell t}}  =  0 \qquad \Longrightarrow
\end{displaymath}
\begin{equation}
 \kappa(\Lambda) \left[ \Lambda^2 -  m_t^{(0)2}(\Lambda)  
\ln \left( \frac{\Lambda^2}{ m_t^{(0)2}(\Lambda)} + 1 \right) \right] 
= \frac{\lambda(\Lambda)}{12} \langle \varphi \rangle_{1\ell t}^2
-\frac{1}{2} M^2(\Lambda) \ ,
\label{1ltgap1}
\end{equation}
where we denoted
\begin{equation}
\kappa(\Lambda) = \frac{ g_t^2(\Lambda) N_{\mbox{\scriptsize c}} }
{16 \pi^2} \ ,  \qquad 
m_t^{(0)}(\Lambda) = 
\frac{g_t(\Lambda) \langle \varphi \rangle_{1\ell t}}{\sqrt{2}} \ .
\label{1ltgapdef}
\end{equation}
In the present framework, we should regard $m_t^{(0)}(\Lambda)$
as the running mass of the top quark at the energy of the upper cut-off
$E=\Lambda$, i.e., the bare mass of the top quark. Later in this Section
we will show that this mass is approximately equal, apart from small
radiative corrections, to the physical mass $m_t^{\mbox{\footnotesize phy}}$.
Furthermore, we can regard, in the present framework, $\langle \varphi 
\rangle_{1\ell t}$ as the actual VEV $v$ ($= 246.2 \mbox{ GeV}$),
apart from small radiative corrections that will be accounted for later.

Note that the $\Lambda^2$-term plays a crucial role in this relation,
similarly as it does also in the $\langle \bar t t \rangle$-condensation
mechanism. Formally, the only difference in (\ref{1ltgap1}) from the
usual leading-$N_{\mbox{\scriptsize c}}$ gap equation of the
TSM-type $\langle  \bar t t \rangle$-condensation
is the presence of the $\lambda(\Lambda)$-terms, due to the
self-interaction of the scalar sector. The ``gap'' equation
(\ref{1ltgap1}) can be regarded as representing a perturbative
influence of the top quark sector on the scalar sector only if
the resulting VEV $\langle \varphi \rangle_{1\ell t}$ is relatively
close to the tree level VEV $\langle \varphi \rangle_0 =
\sqrt{6 M^2(\Lambda)/\lambda(\Lambda)}$. Otherwise, the relation
(\ref{1ltgap1}) should be regarded as an inherently non-perturbative
effect of the top sector on $V_{\mbox{\footnotesize eff}}$.
For example, for $M^2(\Lambda) \approx 0$, i.e., $\langle \varphi
\rangle_0 \ll 246 \mbox{ GeV}$, the effects are highly non-perturbative.
The above ``gap'' equation can be rewritten as
\begin{equation}
2 \kappa(\Lambda) \frac{\Lambda^2}{m^2(\Lambda)} =
\left[ 1 - z_1 \ln \left( z_1^{-1} + 1 \right) \right]^{-1}  \qquad
\left( \ \stackrel{>}{\approx} \ 1 \right) \ ,
\label{1ltgap2}
\end{equation}
\begin{equation}
\mbox{ where we denote:} \quad
m^2(\Lambda) = \frac{\lambda(\Lambda)}{6}\langle \varphi 
\rangle^2_{1\ell t}
 - M^2(\Lambda) \ , \qquad z_1 = \frac{ m_t^{(0)2}(\Lambda)}{\Lambda^2} 
\quad ( \ < \ 1 ) \ .
\label{1ltgapdef2}
\end{equation}
We have $m^2(\Lambda) > 0$, which means that the following relation
between the bare parameters $M^2(\Lambda)$ and $\lambda(\Lambda)$
of the starting scalar potential and the
solution $\langle \varphi \rangle^2_{1\ell t}$ is automatically fulfilled
\begin{equation}
M^2(\Lambda) < \lambda(\Lambda) \langle \varphi \rangle^2_{1\ell t}/6 \ .
\label{1ltgapcon}
\end{equation}
In the case of the tree-level SSB ($M^2(\Lambda) > 0$), we then have by
(\ref{1ltgap1})
\begin{equation}
\langle \varphi \rangle^2_{1 \ell t} - \langle \varphi \rangle^2_0
 = 12 \frac{\kappa(\Lambda)}{\lambda(\Lambda)} \Lambda^2
 \left[ 1-z_1 \ln ( z_1^{-1}+1 ) \right] \ \stackrel{<}{\approx} \
 12 \frac{\kappa(\Lambda)}{\lambda(\Lambda)} \Lambda^2 \ ,
 \label{delphi0}
 \end{equation}
 where $\langle \varphi \rangle_0 =  \sqrt{6 M^2(\Lambda)/
 \lambda(\Lambda)}$ is the VEV at the tree level. We see that the
 1-loop top quark effects, in the case of the tree-level SSB, increase
 the square of the VEV by a term roughly proportional to $\Lambda^2$.

If, on the other hand, $M^2(\Lambda)$ is non-positive, we have no SSB at the
tree level, but we do have it after the inclusion of the top quark effects --
i.e., the quantum-induced SSB according to (\ref{1ltgap1}).

The central question appearing here is: can we obtain any restrictions
on the cut-off $\Lambda$ from the heavy-top-influenced
``gap'' equation (\ref{1ltgap1})? And, what are the resulting masses of 
the Higgs?
The relations (\ref{1ltgap1}) and (\ref{1ltgap2}) can be rewritten
\begin{equation}
\Lambda^2 = \frac{m^2(\Lambda)}{2 \kappa(\Lambda) 
\left( 1-\theta(\Lambda) \right)}
 \ \stackrel{<}{\approx} \ 
\frac{\lambda(\Lambda) \langle \varphi \rangle^2_{1\ell t} }
{12 \kappa(\Lambda) \left( 1-\theta(\Lambda) \right)} \ ,
\label{Lub1}
\end{equation}
\begin{equation}
\mbox{ where} \qquad \theta(\Lambda) = \frac{m_t^{(0)2}(\Lambda)}
{\Lambda^2} \ln \frac{\Lambda^2}{m_t^{(0)2}(\Lambda)} +
{\cal {O}}\left( \frac{m_t^4}{\Lambda^4} \right) \ ,
\label{deldef}
\end{equation}
and the inequality $\stackrel{<}{\approx}$ in (\ref{Lub1}) holds when
either $M^2(\Lambda) \geq 0$, or $M^2(\Lambda)$ is negative
and satisfying $-M^2(\Lambda)= |M^2(\Lambda)| \ll \lambda(\Lambda)
\langle \varphi \rangle^2_{1 \ell t} / 6$. The parameters appearing in
this upper bound are bare parameters -- at the ``running'' energy
$\Lambda$ of the ultimate ultraviolet cut-off of the theory.
We would like to express this upper bound
for the cut-off $\Lambda$ in terms of renormalized parameters, i.e.,
in terms of physical quantities which are, with the exception of the 
physical Higgs mass $M_H$, reasonably well known. Therefore, we devote
the next few paragraphs to the relations between the bare parameters
appearing in (\ref{Lub1}) and the physical parameters, within the present
framework of including only the 1-loop effects of the heavy quark.

The relation between the physical (pole) mass $M_H^{\mbox {\scriptsize
pole}}$ and the corresponding bare coupling $\lambda(\Lambda)$,
as well as the relation
between the physical VEV $\langle \varphi_{\mbox {\scriptsize ren.}}
\rangle = v$ and the bare VEV $\langle \varphi \rangle$,
in the present framework of included 1-loop top quark effects only,
are expressed by means of the following truncated Green function,
corresponding to the diagram of Fig.~2a
\begin{equation}
-i \Sigma_{HH}( q^2 )  = -i \Sigma_{HH}^{tt} ( q^2 )
= \left( \frac{g_t(\Lambda)}{\sqrt{2}} \right)^2
N_{\mbox {\footnotesize c}} \int \frac{d^4 k}{(2\pi)^4}
tr_f \left[ 
\frac{i}{\left( {k \llap /} - m_t^{(0)}(\Lambda) \right)}
\frac{i}{\left( {k \llap /} + {q \llap /} - m_t^{(0)}(\Lambda) \right)}
\right] \ .
\label{Sigmatt1}
\end{equation}
These relations are
\begin{eqnarray}
\left( M_H^{\mbox{\scriptsize pole}} \right)^2 & = &
\frac{ d^2 V^{(0)} }{d \varphi^2} {\Big|}_
{\varphi= \langle \varphi \rangle_{1\ell t}} + \Sigma_{HH}\left(
q^2 = \left( M_H^{\mbox{\scriptsize pole}} \right)^2 \right)
\nonumber\\
& = & 
\frac{d^2 V_{\mbox{\footnotesize eff}}^{(0+1\ell t)} }
{d \varphi^2} {\Big|}_
{\varphi= \langle \varphi \rangle_{1\ell t}} + \Sigma_{HH}\left(
q^2 = \left( M_H^{\mbox{\scriptsize pole}} \right)^2 \right)
- \Sigma_{HH}\left( q^2 = 0 \right) \ ,
\label{MHpole1}
\end{eqnarray}
\begin{equation}
Z_{\varphi} \varphi^2 = \varphi^2_{\mbox{\footnotesize ren.}} \ , \qquad
\mbox{where: } Z_{\varphi}= \left[ 1 - \frac{d \Sigma_{HH}(q^2)}{d q^2}
{\Big|}_{q^2=(M_H^{pole})^2} \right] = 1 + \delta Z_{\varphi} \ .
\label{phiren}
\end{equation}
In particular, the renormalized VEV $v = \langle \varphi_{\mbox
{\footnotesize ren.}} \rangle$ ($= 246.2 \mbox{ GeV}$) is related 
to the bare VEV $\langle \varphi \rangle$ ($=\langle \varphi \rangle_{1\ell t}$
in this Section) by
\begin{equation}
\langle \varphi \rangle^2 = \left( 1 - \delta Z_{\varphi} \right) v^2 \ .
\label{VEVren}
\end{equation}
These relations can be obtained by simply summing up in the Higgs
propagator the $t \bar t$-loops of Fig.~2a in the leading-log
approximation (geometrical series), starting with the ``bare'' Higgs
propagator with the mass equal to the $d^2 V^{(0)}/d \varphi^2$
evaluated at the {\it corrected}~\footnote{
It is at this corrected VEV that the linear term ($\propto
H = \varphi - \langle \varphi \rangle$) of the tree level
potential $V^{(0)}$ is canceled by the 1-loop top quark tadpole.}
VEV $\langle \varphi \rangle_{1\ell t}$.
$\Sigma_{HH}^{tt}$ can be calculated directly, by performing first
the Wick rotation in the Euclidean space. Then we impose
again the spherical cut-off $\Lambda$ on the Euclidean top quark 
momentum $\bar k$ and end up with the following result
\begin{equation}
\Sigma_{HH}^{tt}\left( q^2 \right) =  -2 \kappa(\Lambda)
{\Big \{}  \Lambda^2 +\left[ \frac{q^2}{2} - 
3 m_t^{(0)2}(\Lambda) \right]
 \ln \left( \frac{\Lambda^2}{ m_t^{(0)2}(\Lambda) } \right)
+ {\cal {O}}(q^2,m_t^2) {\Big \} } \ .
\label{Sigmatt2}
\end{equation}
The cut-off independent part can also be calculated explicitly
for Euclidean $\bar q^2 = - q^2 > 0$, and then it can be analytically
continued into the physical region $q^2 > 0$. The value of this part,
however, depends on the choice of the regularization (cut-off procedure).
We will ignore these terms; they are smaller than the 
$\ln (\Lambda^2/m_t^2)$-term even in the case of low $\Lambda \sim 
\mbox{1 TeV}$ -- typically by a factor of 3 or more. This approximation
will not affect our results appreciably, because even the 
$\ln \Lambda$-terms from the above $\Sigma_{HH}^{tt}$ will contribute
only a relative correction of less than 10 percent
to the physical parameters of our concern.

Inserting (\ref{Sigmatt2}) into (\ref{phiren}), we obtain the
$\varphi$-renormalization parameter $\delta Z_{\varphi}$
\begin{equation}
\delta Z_{\varphi} =  \kappa(\Lambda)
{\Big \{} \ln \left[ \frac{\Lambda^2}{m_t^{(0)2}(\Lambda) }
\right] + {\cal {O}}(\Lambda^0) {\Big \}} \ .
\label{Zphi}
\end{equation}
Inserting (\ref{Sigmatt2}) into (\ref{MHpole1}), the pole mass of the
Higgs in the present framework acquires the form
\begin{equation} \left( M_H^{\mbox{\scriptsize pole}} \right)^2 =
\frac{d^2 V_{\mbox{\footnotesize eff}}
^{(0+1\ell t)} }{d \varphi^2} {\Big |}
_{\varphi = \langle \varphi \rangle_{1\ell t} } 
{\Big \{ } 1 - \kappa(\Lambda) \left[ \ln \left( \frac{\Lambda^2}
{(m_t^{(0)}(\Lambda))^2} \right) + {\cal{O}}(\Lambda^0) \right] 
\cdots {\Big \} } \ ,
\label{MHpole2}
\end{equation}
where ($\ldots$) represent higher powers of $\kappa \ln(\Lambda/m_t)$.
These terms are small, because for $m_t \approx 180 \mbox{ GeV}$ 
we have $\kappa \approx 2 \cdot 10^{-2}$ (-- if we for a moment
ignore the differences between the physical and bare quantities).
The second derivative appearing in the above relation can be
directly calculated from (\ref{Vtree})-(\ref{V1l}) and
the ``gap'' equation (\ref{1ltgap1})
\begin{equation}
 \frac{d^2 V_{\mbox{\footnotesize eff}}
^{(0+1\ell t)} }{d \varphi^2} {\Big |}
_{\varphi = \langle \varphi \rangle_{1\ell t} } 
= {\Big \{ } \frac{\lambda(\Lambda)}{3}
 + 2 \kappa(\Lambda) g_t^2(\Lambda) 
\left[ \ln\ \frac{\Lambda^2}{ m_t^{(0)2}(\Lambda) } - 1 
+ {\cal {O}}\left( \frac{m_t^2}{\Lambda^2} \right)
\right] {\Big \}} \langle \varphi\rangle^2_{1\ell t} \ .
\label{MHpole2rf}
\end{equation}
The formulas (\ref{MHpole2}) and (\ref{MHpole2rf}) give us a connection
between $M_H^{\mbox{\footnotesize pole}}$ and the bare coupling
$\lambda(\Lambda)$ in terms of the cut-off $\Lambda$
and the bare values for the 
Yukawa coupling and the VEV (at the ``running''
energy of the upper cut-off $E \approx \Lambda$). We would like to
have a corresponding relation in terms of renormalized quantities
$g_t^{\mbox{\scriptsize ren.}}$ and $v$, because then 
we can find the connection
between the $\Lambda$-upper bound (\ref{Lub1}) 
and the physical mass $M_H^{\mbox{\footnotesize pole}}$ in terms of 
these well-known renormalized quantities. 

Within the present framework, we have the following
relation for the Yukawa coupling
\begin{equation}
g_t(\Lambda) = Z_{gt}^{-1/2} g_t^{\mbox{\scriptsize ren.}}  \ ,
\qquad \left( \Rightarrow \kappa(\Lambda) = Z_{gt}^{-1} \kappa_{\mbox
{\scriptsize ren.}} \right) \ ,
\label{Yukren}
\end{equation}
where the renormalization constant $Z_{gt}$ can most easily be
obtained by considering the renormalization group equation (RGE) for
the $g_t$, running it down from $E \approx \Lambda$ to $E \approx
m_t$
\begin{equation}
Z_{gt} = 1 + \delta Z_{gt} \ ; \qquad \delta Z_{gt} = 
- \frac{3}{2} \delta Z_{\varphi} + 2 \delta Z_{gl} \ .
\label{Zgl1}
\end{equation}
Here, we use $\delta Z_{\varphi}$ of (\ref{Zphi}), and
the gluonic renormalization effects are parametrized by $\delta Z_{gl}$
\begin{equation}
\delta Z_{gl} =  \frac{\alpha_s}{\pi} \left[ \ln  \frac{\Lambda^2}
{ m_t^{(0)2} \left( \Lambda \right) }
+ {\cal {O}}(\Lambda^0)  \right] \ .
\label{Zgl2}
\end{equation}
We took here into account only the numerically dominant
contributions to the RGE-running of $g_t$: the QCD contribution
$2 \delta Z_{gl}$ ($\propto \alpha_s/\pi$, where $\alpha_s \approx
\alpha_s(E \stackrel{>}{\sim} m_t) \approx 0.10$), 
and the Yukawa contribution $-3 \delta Z_{\varphi}/2$
($\propto \kappa(\Lambda)$, where
$\kappa(\Lambda) \sim \kappa_{\mbox{\scriptsize ren.}} =
2.03 \cdot 10^{-2}$, for $m_t^{\mbox{\footnotesize phy}} = 180 
\mbox{ GeV}$). Combining (\ref{Yukren})--(\ref{Zgl2}) with the
known effect of the ``running'' of the VEV between $m_t$ and $\Lambda$ 
(\ref{VEVren}),
we also find the desired connection between
the bare and the renormalized mass of the top quark 
\begin{equation}
m_t^{(0)}(\Lambda) =  \left( \frac{ g_t(\Lambda) \langle \varphi \rangle
_{1\ell t} }{\sqrt{2}} \right) = \left(1 - \frac{1}{2} \delta Z_{gt}
- \frac{1}{2} \delta Z_{\varphi} \right) m_t^{\mbox{\footnotesize phy}}
= \left( 1 - \delta Z_{gl} + \frac{1}{4}
\delta Z_{\varphi} \right) m_t^{\mbox{\footnotesize phy}} \ .
\label{mtren}
\end{equation}
One may ask the question whether $\delta Z_{gt}$ in
(\ref{Zgl1})--(\ref{mtren}) is really free of any $\Lambda^2$-terms,
especially since the RGE for $g_t$ would ignore any such terms.
This is equivalent to the question whether the combination
$(\delta Z_{gt} + \delta Z_{\varphi})$ in (\ref{mtren}) is free
of any $\Lambda^2$-terms, since $\delta Z_{\varphi}$ was shown
explicitly to satisfy this condition (cf.~(\ref{phiren})--(\ref{Zphi})),
in the present framework where we ignore the 1-loop
scalar self-interaction effects. Indeed, the combination
$(\delta Z_{gt} + \delta Z_{\varphi})$ in (\ref{mtren})
can also be obtained independently, by calculating the 1-loop contributions
of the scalar and the quark sectors (taking $g_t$ as the only nonzero
Yukawa coupling) to the propagator of the top quark and finding the
corresponding change of the pole mass $\delta m_t =
m_t^{\mbox{\footnotesize phy}} - m_t^{(0)}(\Lambda)$. It is crucial,
however, to take in the tree-level propagator the mass equal to
$m_t^{(0)}(\Lambda)$ of (\ref{1ltgapdef}). In the resulting $\delta
m_t$ {\it no} $\Lambda^2$-terms appear, i.e., the tadpole diagrams (Fig.~3)
do not contribute, as ensured by the ``gap'' equation (\ref{1ltgap1}).
As a matter of fact, it can be shown that the condition of the cancelation
of the tadpole diagrams of Fig.~3 is equivalent to this ``gap''
equation. All $\Lambda^2$-terms are contained in the bare mass $m_t^{(0)}
(\Lambda)$.

Now, we can express the bare quantities appearing in the
preceding equations through the following renormalized known quantities
\begin{eqnarray}
m_t^{\mbox{\footnotesize phy}}& =& 180 \mbox{ GeV} \ , \qquad
\langle \varphi_{\mbox{\footnotesize ren.}} \rangle = v = 246.2 \mbox{ GeV} \ ,
\nonumber\\
g_t^{\mbox{\scriptsize ren.}}& =& 
\frac{ m_t^{\mbox{\footnotesize phy}} \sqrt{2} }{v} = 1.034 \ , \qquad 
\kappa_{\mbox{\scriptsize ren.}}= 
\frac{\left(g_t^{\mbox{\scriptsize ren.}}\right)^2 N_{\mbox{\scriptsize c}} }
{16 \pi^2} =2.03 \cdot 10^{-2} \ .
\label{renpar}
\end{eqnarray}
For simplicity, we will omit from now on any superscripts
or subscripts ``phy'', ``pole'', ``ren.'' for the parameters $m_t$,
$g_t$, $\kappa$ and $M_H$. Unless otherwise stated,
these parameters will be the physical ones.
On the other hand, we continue to denote by $\varphi$ the bare scalar
field $\varphi^{(\Lambda)}$.

Furthermore, in the spirit of perturbation, we can replace in the
relations (\ref{Zphi})-(\ref{MHpole2}) and (\ref{Zgl2}) the bare quantities
$\kappa(\Lambda)$ and $m_t^{(0)}(\Lambda)$ by the corresponding
renormalized ones (\ref{renpar})
\begin{equation}
\delta Z_{\varphi} = \kappa \left[ \ln \frac{\Lambda^2}{m^2_t} +
{\cal {O}} \left( \Lambda^0 \right) \right] \ ,
\quad \delta Z_{gl} =  \frac{\alpha_s}{\pi} \left[ \ln  \frac{\Lambda^2}
{ m^2_t} + {\cal {O}}(\Lambda^0)  \right] \ ,
\quad ( \delta Z_{gt} = 
-\frac{3}{2} \delta Z_{\varphi} + 2 \delta Z_{gl} ) \ .
\label{delZs}
\end{equation}
Relations (\ref{VEVren})-(\ref{delZs}) allow us now
to express the physical (pole) mass $M_H$ of
(\ref{MHpole2}) in terms of the known renormalized parameters (\ref{renpar})
and of the bare coupling parameter $\lambda(\Lambda)$ and the cut-off
$\Lambda$
\begin{equation}
M^2_H =
\frac{\lambda(\Lambda) v^2}{3} {\Big \{ }
1 + \delta Z_{\varphi} \left[ -2 + \frac{12 m_t^2}{\lambda(\Lambda) v^2} \right]
+ {\cal {O}} \left( \frac{\kappa g^2_t}{\lambda(\Lambda)} \right)
{\Big \} } \ .
\label{MHpole3}
\end{equation}
We can now finally express the upper bound (\ref{Lub1}) for the cut-off 
$\Lambda$ in terms of the renormalized parameters $M_H$, $m_t$, $\kappa$
with the help of relation (\ref{MHpole3})
and ``renormalization'' relations (\ref{VEVren}),
(\ref{Yukren})--(\ref{Zgl1}), (\ref{delZs})
\begin{equation}
\Lambda^2 \ \stackrel{<}{\approx} \ \frac{M^2_H}{4 \kappa} 
\frac{1}{\left( 1 - \theta(\Lambda) \right)}
\left[ 1 + 2 \delta Z_{gl} - \delta Z_{\varphi}
\left( \frac{1}{2} + \frac{4 m^2_t}{M^2_H} \right) 
+ {\cal {O}} \left( \kappa, \frac{\alpha_s}{\pi} \right) \right] \ ,
\label{Lub2}
\end{equation}
\begin{equation}
\mbox{ when: } \qquad M^2(\Lambda) \ \geq \ 0 \ ,
\quad \mbox{or } \quad -M^2(\Lambda)= |M^2(\Lambda)| \ \ll \ 
\frac{\lambda(\Lambda) \langle \varphi \rangle^2_{1 \ell t}}{ 6 }
\ \left( \approx \frac{\lambda(\Lambda) v^2}{ 6 } \right) \ .
\label{Lubco}
\end{equation}
Here, $\theta(\Lambda)$ is the small parameter defined through
(\ref{deldef}) and (\ref{1ltgapdef}).
For example, for $\Lambda \sim \mbox{1 TeV}$ we have
$\theta(\Lambda) \approx 0.1$. Within the framework of this section,
the bare mass $m_t^{(0)}(\Lambda)$ is close to the physical mass
$m_t =180 \mbox{ GeV}$. Therefore
\begin{equation}
\theta(\Lambda) = \frac{m^2_t}{\Lambda^2} \ln \frac{\Lambda^2}{m^2_t}
\left[1 + {\cal {O}}\left(\delta Z_{gl,\varphi}\right) \right] \ .
\label{deldef2}
\end{equation}
Relation (\ref{Lub2}) gives us, after one or two iterations,
an upper bound on the cut-off $\Lambda$ as a function of the
(physical) Higgs mass $M_H^2$. This upper bound arised
as a consequence of the (non-perturbative) effect of the heavy top quark 
on the scalar sector. 
Alternatively, we can simply express this upper bound with the bare 
coupling $\lambda(\Lambda)$ instead of $M^2_H$, using (\ref{Lub1})
and the renormalization conditions (\ref{VEVren}) and 
(\ref{Yukren})--(\ref{Zgl1})
\begin{equation}
\Lambda^2 \ \stackrel{<}{\approx} \ \frac{\lambda(\Lambda) v^2}
{12 \kappa} \frac{1}{\left(1-\theta(\Lambda) \right) }
\left[ 1 + \delta Z_{gt} - \delta Z_{\varphi} \right] \ ,
\label{Lub2a}
\end{equation}
where the conditions for the inequality are those in (\ref{Lubco}).
We note that for $|M(\Lambda)| \ll  v \sqrt{\lambda(\Lambda)/6}$
($\approx M_H/\sqrt{2}$), the actual ultraviolet cut-off Lambda becomes 
approximately equal to the upper bound in (\ref{Lub2}), (\ref{Lub2a}) .

Some numerical results of this relation
($\lambda(\Lambda)$ vs.~$M_H$ vs.~$\Lambda^{\mbox{\scriptsize u.b.}}$)
are given in the first three columns of Table 1. We note that we obtain 
rather low upper bounds 
$\Lambda^{\mbox{\scriptsize u.b.}} \stackrel{<}{\approx} 1 \mbox{ TeV}$.

These results basically survive even in the case when $-M^2(\Lambda) =
|M^2(\Lambda)| = {\cal {O}}(\lambda v^2)$ ($={\cal {O}}(M^2_H)$ by
(\ref{MHpole3})). The values of $M_H$ as a function of $\lambda(\Lambda)$
and $\Lambda$ remain unaffected then, as seen by formula (\ref{MHpole3}).
This is so because any $M^2(\Lambda)$-dependence there has been eliminated
by the use of the ``gap'' equation (\ref{1ltgap1}). The expression on the
r.h.s.~of (\ref{Lub2a}), however, is modified in such a case by the
replacement: $\lambda(\Lambda) \langle \varphi \rangle^2_{1\ell t}/6
\mapsto m^2(\Lambda)$, as seen from (\ref{Lub1}) and (\ref{1ltgapdef2}).
This result in an additional factor $k(\Lambda)$ of order 1 on the
r.h.s.~of (\ref{Lub2a})
\begin{equation}
\Lambda^2  = \frac{k(\Lambda) \lambda(\Lambda) v^2}
{12 \kappa} \frac{1}{\left(1-\theta(\Lambda) \right) }
\left[ 1 + \delta Z_{gt} - \delta Z_{\varphi} \right] \ ,
\label{Lub2b}
\end{equation}
\begin{equation}
\mbox{where: } \quad k(\Lambda)  =  \left[ 1 +
\frac{6 |M^2(\Lambda)|}{\lambda(\Lambda) v^2}\left( 1 + \delta Z_{\varphi}
\right) \right] \ , \qquad
\mbox{for: } \
-M^2(\Lambda) = |M^2(\Lambda)| = {\cal {O}}(\lambda(\Lambda) v^2) \ .
\label{Lubco1b}
\end{equation}
Therefore, $\Lambda = {\cal {O}}(1 \mbox{ TeV})$ also in this case.

We stress that all these results are in the framework of the present Section
where the 1-loop contributions of the scalar self-interaction and of the 
electroweak gauge bosons to $V_{\mbox{\footnotesize eff}}$ have been ignored.
In the next Section, we will include these contributions.

\section{Inclusion of 1-loop scalar and gauge bosonic contributions}

It is straightforward to obtain from (\ref{Vtree}) and (\ref{V1l})
an explicit expression for the effective potential with the leading
heavy top quark contributions included
\begin{equation}
V_{\mbox{\footnotesize eff}}^{(0+1\ell t)} \left(\varphi^2; \Lambda
\right)
= - \frac{1}{2} M_0^2 \varphi^2 + \frac{1}{4!} \lambda_0 \varphi^4
  - \frac{1}{4} \kappa(\Lambda) g_t^2(\Lambda) \varphi^4 \ln
 \frac{\varphi^2}{\langle \varphi \rangle^2_{1\ell t} }
+ {\cal {O}}\left( \Lambda^{-2} \right) \ ,
\label{V1lt2}
\end{equation}
where we denoted
\begin{eqnarray}
M_0^2 & =&  M^2(\Lambda) + 2 \kappa(\Lambda) \Lambda^2 =
\left[ \frac{\lambda(\Lambda)}{6} + g^2_t \delta Z_{\varphi} \left(
1 + {\cal {O}}\left( 1 / \ln \frac{\Lambda^2}{m^2_t} \right) \right) \right]
\langle \varphi \rangle^2_{1\ell t} \ , \nonumber\\
\lambda_0 & = & \lambda(\Lambda) + 6 g^2_t \delta Z_{\varphi} \left(
1 + {\cal {O}}\left( 1 / \ln \frac{\Lambda^2}{m^2_t} \right) \right) \ .
\label{V1ltdef}
\end{eqnarray}
Here, $\delta Z_{\varphi}$ is the expression written in (\ref{Zphi}),
or in an approximate form in (\ref{delZs}).
Strictly speaking, the Yukawa coupling in (\ref{V1ltdef}) is the bare
one ($g_t(\Lambda)$), and $\delta Z_{\varphi}$ is defined by (\ref{phiren})
(with $\Sigma_{HH} \mapsto \Sigma_{HH}^{tt}$) and (\ref{Zphi}). It contains
the ``bare'' mass $m_t^{(0)}(\Lambda)$ defined in (\ref{1ltgapdef}).
However, since it will turn out that $\langle \varphi \rangle^2 \sim
\langle \varphi \rangle^2_{1\ell t}$, and since we will neglect the terms
${\cal {O}}(1/\ln (\Lambda^2/m^2_t))$ in (\ref{V1ltdef}), we can use
instead there the $\delta Z_{\varphi}$ given by (\ref{delZs}) in terms of
physical quantities. The errors due to the replacement $g_t(\Lambda)
\mapsto g_t$ in (\ref{V1ltdef}) are even smaller
(cf.~(\ref{Yukren})--(\ref{Zgl2}) and (\ref{delZs})).
The expression on the r.h.s.~of the first line in (\ref{V1ltdef}) was
obtained from the ``gap'' equation (\ref{1ltgap1}). 

The effective potential (\ref{V1lt2}) is now used as the starting point
(i.e., ``tree level'') to calculate the perturbative 1-loop corrections
to it coming from the scalar sector itself and from the sector of the
electroweak gauge bosons. This can be done in a straightforward way
by using, for example, the standard path integral approach~\cite{huang}. 
This leads us to the following corrections to $V_{\mbox{\footnotesize
eff}}$:
\begin{eqnarray}
\delta V^{(1\ell sc)}_{\mbox{\footnotesize eff}}\left( \varphi^2;
\Lambda \right) 
& = &  \frac{1}{64 \pi^2} {\Big \{ } 2 \Lambda^2 \left[ \left(
\frac{\lambda_0}{2} \varphi^2 - M_0^2 \right) + 3 \left(
\frac{\lambda_0}{6} \varphi^2 - M_0^2 \right) + {\cal {O}}\left(
\kappa g^2_t \varphi^2 \right) \right] 
\nonumber\\
&&- \left( \ln \frac{\Lambda^2}{m_t^2} \right) \left[ 
\left( \frac{\lambda_0}{2} \varphi^2 - M_0^2  + 
{\cal {O}}\left(\kappa g^2_t \varphi^2 \right) \right)^2 + 3 \left(
\frac{\lambda_0}{6} \varphi^2 - M_0^2 +
{\cal {O}}\left(\kappa g^2_t \varphi^2 \right) \right)^2 \right] 
\nonumber\\
&&+ {\cal {O}}\left( (\lambda_0 \varphi^2/3)^2 \right) {\Big \}} \ ,
\label{V1lsc}
\end{eqnarray}
\begin{equation}
\delta V^{(1\ell g b)}_{\mbox{\footnotesize eff}} \left( \varphi^2;
\Lambda \right)
 =  \frac{3}{32 \pi^2 } {\Big \{} \int_0^{\Lambda^2}
d{\bar q}^2 {\bar q}^2 \ln \left[ \frac{ {\bar q}^2 + M^2_Z \varphi^2/v^2}
{{\bar q}^2} \right] + 
2  \int_0^{\Lambda^2}
d{\bar q}^2 {\bar q}^2 \ln \left[ \frac{ {\bar q}^2 + M^2_W \varphi^2/v^2}
{{\bar q}^2} \right] {\Big \}} \ .
\label{V1lgb}
\end{equation}
The latter expression was calculated in the Landau gauge
($\xi \to \infty$), since only in this gauge the coupling
of ghosts to scalars is zero and we don't have any 1-loop
contribution of ghosts to $V_{\mbox{\footnotesize eff}}$.
Furthermore, we replaced in (\ref{V1lgb}) the combinations
$g^2(\Lambda)+g^{\prime 2}(\Lambda)$ and $g^2(\Lambda)$ of the
bare electroweak coupling parameters by their tree level
approximations $4 M^2_Z/v^2$ and $4 M^2_W/2$, respectively. The
error arising from this replacement will be very minor
(for $\Lambda = {\cal {O}}(1 \mbox{ TeV})$).

We note that the expression (\ref{V1lsc}), representing the 1-loop contribution
of the Higgs and of the three Goldstones to the effective potential 
$V_{\mbox{\footnotesize eff}}$, should be regarded as a {\it perturbative} 
correction to $V_{\mbox{\footnotesize eff}}^{(0+1 \ell t)}$. This is so 
because (\ref{V1lsc}) represents an effect of the scalar sector on
{\it itself}, and the higher loop contributions of such effects would
be proportional to the correspondingly higher powers in the modified scalar
self-interaction parameter $\lambda_0$. It would be inconsistent
to equate in such a series the terms corresponding to different powers
of $\lambda_0$. Hence, we are
able to make reasonable predictions about these corrections only as
long as the scalar sector is not too strongly self-interacting, i.e.,
as long as these corrections do not drastically change, through its
contributions, the VEV $\langle \varphi \rangle^2_{1\ell t}$ and
the upper bound $(\Lambda^{\mbox{\scriptsize u.b.}}_{1\ell t})^2$ 
of the previous Section --
say, by not more than 50 percent. We will see soon that this
will restrict our predictions to a region of Higgs masses
$M_H \stackrel{<}{\approx} 250 \mbox{ GeV}$. 

On the other hand, 
the 1-loop corrections (\ref{V1lgb}) to $V_{\mbox{\footnotesize eff}}$
from the sector of the electroweak gauge bosons can, in principle, 
be regarded as valid also in the non-perturbative
region -- i.e., if the gauge boson masses and the resulting corrections 
to the ``gap'' equation were large. However, for the
experimentally well-known values of $M_Z$ and $M_W$, these contributions
will turn out to be quite small, and we will treat them as perturbative
corrections.

The ``gap'' equation (\ref{1ltgap1}), which determines the bare VEV when
the corrections (\ref{V1lsc}) and (\ref{V1lgb}) to $V_{\mbox
{\footnotesize eff}}$ are ignored, is now modified perturbatively
\begin{displaymath}
\frac{d}{d \varphi^2} \left( 
V^{(0+1 \ell t)}_{\mbox{\footnotesize eff}} 
+\delta V^{(1\ell sc)}_{\mbox{\footnotesize eff}}
+\delta V^{(1\ell g b)}_{\mbox{\footnotesize eff}} \right) {\Big |}_
{\varphi = \langle \varphi \rangle} = 0 \qquad \Longrightarrow
\end{displaymath}
\begin{equation}
\delta \langle \varphi \rangle^2 
\frac{ d^2 V_{\mbox{\footnotesize eff}}^{(0+1\ell t)} }
{d \left( \varphi^2 \right)^2} {\Bigg |}_
{\varphi^2 = \langle \varphi \rangle^2_{1\ell t}}
+ \frac{ d \left( \delta V^{(1\ell sc)}_{\mbox{\footnotesize eff}} \right) }
{d \varphi^2 } {\Bigg |}_{\varphi^2 = \langle \varphi \rangle^2_{1\ell t}}
+ \frac{ d \left( \delta V^{(1\ell gb)}_{\mbox{\footnotesize eff}} \right) }
{d \varphi^2 } {\Bigg |}_{\varphi^2 \approx v^2} \approx 0 \ ,
\label{1lgap}
\end{equation}
where we denoted
\begin{equation}
\delta \langle \varphi \rangle^2 = \langle \varphi \rangle^2 -
\langle \varphi \rangle^2_{1\ell t} \ .
\label{delphidef}
\end{equation}
We mean by $\langle \varphi \rangle$ the corrected VEV of the
(bare) scalar field $\varphi = \varphi^{(\Lambda)}$. In the perturbative
approach, we demand that the correction (\ref{delphidef}) be
relatively small. That's why we included in the corrected ``gap''
equation (\ref{1lgap}) only the first two terms in the Taylor expansion
of the derivative
$d V_{\mbox{\footnotesize eff}}^{(0+1\ell t)}/d{\varphi}^2$
around the value $\varphi^2 =
\langle \varphi \rangle^2_{1\ell t}$. Note that the latter value,
by definition, satisfies the (0+1-loop top) ``gap'' equation 
(\ref{1ltgap1}). On the other hand, the derivative of the 
(smaller) corrective potentials
$\delta V^{(1\ell sc)}_{\mbox{\footnotesize eff}}$ and
$\delta V^{(1\ell g b)}_{\mbox{\footnotesize eff}}$ in
(\ref{1lgap}) are evaluated, according to convenience, at one of
the following similar values of VEV:
$\langle \varphi \rangle^2_{1\ell t}$
and  $\langle \varphi_{\mbox{\footnotesize ren.}} \rangle^2 = v^2$,
i.e., here we take just the first term in Taylor expansions.

Explicit calculation of (\ref{1lgap}) then leads to the following
expression for the VEV correction $\delta \langle \varphi \rangle^2$:
\begin{eqnarray}
\delta \langle \varphi \rangle^2 & = &   
- \frac{3}{8 \pi^2} \Lambda^2 \left[ 1 
+ {\cal {O}}\left(\frac{\kappa g^2_t}{\lambda_0} \right) 
+ {\cal {O}}\left( \theta(\Lambda) \right) \right] +
\nonumber\\
&&-\frac{9}{8 \pi^2}\frac{1}{\lambda_0} \Lambda^2
\left( \frac{M_Z^2+2 M_W^2}{v^2} \right) \left[ 1 + {\cal {O}}\left(
\frac{M_Z^2}{\Lambda^2} \ln \frac{\Lambda^2}{M_Z^2} \right) 
+ {\cal {O}}\left(\frac{\kappa g^2_t}{\lambda_0} \right) \right] \ .
\label{delVEV}
\end{eqnarray}
We note that the (0+1-loop top) ``gap'' equation 
(\ref{1ltgap1}) defines $\langle \varphi \rangle^2_{1\ell t}$
in the first place. From there we obtained, for the case of $M^2(\Lambda) 
\geq 0$ or $-M^2(\Lambda) = |M^2(\Lambda)| \ll \lambda(\Lambda)
\langle \varphi \rangle^2_{1 \ell t}/6$,
the upper bound (\ref{Lub1}) for $\Lambda$, expressed in
terms of the bare VEV $\langle \varphi \rangle^2_{1\ell t}$. This
upper bound is, of course, still valid in our case. However,
the bare VEV $\langle \varphi \rangle^2_{1\ell t}$ in (\ref{Lub1})
cannot be regarded in the framework of the present Section as the
actual bare VEV. We express $\langle \varphi \rangle^2_{1\ell t}$ in
(\ref{Lub1}) as the difference between the actual
(i.e., corrected) bare VEV $\langle \varphi \rangle^2$ 
and the scalar and gauge bosonic corrections $\delta \langle \varphi \rangle^2$
of (\ref{delVEV}).
Thus we obtain, for the case of $M^2(\Lambda) 
\geq 0$ or $-M^2(\Lambda) = |M^2(\Lambda)| \ll \lambda(\Lambda) v^2$,
the following upper bound for the ultraviolet cut-off $\Lambda$, 
now modified in comparison to (\ref{Lub2a})
perturbatively by the 1-loop scalar and 1-loop gauge bosonic contributions 
\begin{eqnarray}
\Lambda^2 & \stackrel{<}{\approx} &
\frac{\lambda(\Lambda) v^2}{12 \kappa} \frac{1}
{\left( 1 - \theta(\Lambda) \right)} \left[1 + \delta Z^{\prime}_{gt}
- \delta Z^{\prime}_{\varphi} \right] 
{\Bigg \{} 1 +
\frac{\lambda(\Lambda)}{32 \pi^2 \kappa}\left[1 + 
{\cal {O}}\left(\frac{\kappa g^2_t}{\lambda(\Lambda)}\right)+
{\cal {O}}\left(\theta(\Lambda) \right)
+ {\cal {O}}\left( \delta Z_{gt,\varphi} \right)
\right] 
\nonumber\\
&&+\frac{3}{32 \pi^2 \kappa}\frac{\lambda(\Lambda)}{\lambda_0}
\left(\frac{M_Z^2+2 M_W^2}{v^2}\right)
\left[ 1 + {\cal {O}}\left(\frac{M_Z^2}{\Lambda^2}\ln \frac{\Lambda^2}
{M_Z^2} \right) + 
{\cal {O}}\left(\frac{\kappa g^2_t}{\lambda(\Lambda)}\right) +
+ {\cal {O}}\left( \delta Z_{gt,\varphi} \right)
\right] {\Bigg \}} \ ,
\label{Lub3}
\end{eqnarray}
The parameter $\lambda_0$ appearing above was
defined in (\ref{V1ltdef}) in terms of the bare
parameter $\lambda(\Lambda)$ and the cut-off $\Lambda$. 
The parameters $\delta Z^{\prime}_{\varphi}$ and $\delta Z^{\prime}_{gt}$
are perturbative renormalization effects for $g_t$ and $\varphi$ 
\begin{equation}
\langle \varphi \rangle^2 =
\left( 1 - \delta Z^{\prime}_{\varphi} \right) v^2 \ , \quad
g^2_t(\Lambda) =  \left( 1 - \delta Z^{\prime}_{gt} \right) g^2_t \ ,
\quad \left( \Rightarrow \kappa(\Lambda) =  (1-\delta Z^{\prime}_{gt}) )
\kappa \ , \right) \ .
\label{delZs2}
\end{equation}
They are similar to the expressions $\delta Z_{\varphi}$ and $\delta Z_{gt}$
of the previous Section (cf.~eqs.~(\ref{Zphi}) and 
(\ref{Zgl1})--(\ref{Zgl2})), the only difference being the modifications
by contributions of the electroweak
gauge bosons (see eqs.~(\ref{Zprimephi}) and (\ref{Zprimegt}) and
the discussion below).
The small parameter $\theta(\Lambda)$ appearing in (\ref{Lub3})
was defined through (\ref{deldef}) and (\ref{1ltgapdef}). Therefore,
it can be written now in terms of $\Lambda$ and the physical
parameter $m_t$
\begin{equation}
\theta(\Lambda) =
\frac{m^2_t}{\Lambda^2} \left[ 1 +
\frac{ |\delta \langle \varphi \rangle^2 | }{v^2} \right]
 \ln \left[  \frac{\Lambda^2}{m^2_t \left( 1 +
  |\delta \langle \varphi \rangle^2 |/v^2 \right) } \right]
\left[ 1 + {\cal {O}}\left( \delta Z^{\prime}_{\varphi, gt} \right) \right] \ ,
\label{deldef3}
\end{equation}
where $\delta\langle \varphi \rangle^2 = -|\delta\langle \varphi \rangle^2|$
as a function of $\Lambda$
is given in (\ref{delVEV}). Note that $|\delta \langle \varphi \rangle^2|/
v^2 \stackrel{<}{\sim} 1$, because we restrict ourselves to perturbative
effects of the scalar and gauge bosonic sector.
All the other
parameters appearing in (\ref{Lub3}) are well known (cf.~(\ref{renpar}),
and: $M_W \approx 80.2 \mbox{ GeV}$, $M_Z \approx 91.2 \mbox{ GeV}$).
Relation (\ref{Lub3}) is now the modified version of relation
(\ref{Lub2a}), expressing the upper bound of the ultraviolet cut-off
with the bare self-coupling parameter $\lambda(\Lambda)$. 
In this relation, the second term in the curly brackets represents
the modification due to the 1-loop scalar self-interaction contributions,
and the third the modification due to the 1-loop gauge bosonic
contributions to the effective potential. Strictly speaking, we obtain
an infinite geometric series of such terms; however, in the spirit of
perturbation, we ignore all the terms of higher powers in
$\lambda(\Lambda)/(32 \pi^2 \kappa)$ and $M^2_{g.b.}/(32 \pi^2 \kappa v^2)$.

As already indicated in the previous paragraph,
the inclusion of the 1-loop scalar and
gauge bosonic contributions to the ``gap'' equation requires, for reasons 
of consistency, that we also modify the parameter $\delta Z_{\varphi}$
of eqs.~(\ref{phiren}) and (\ref{Zphi}): $\delta Z_{\varphi} \mapsto
\delta Z^{\prime}_{\varphi}$. The modification
is caused by the 1-loop contributions
of scalars and gauge bosons to the derivative $d\Sigma_{HH}(q^2)/dq^2$.
It turns out that the scalar self-interactions alone do not contribute to
this derivative because they give $q^2$-independent contribution to
the $\Lambda$-dependent part of $\Sigma_{HH}$. The electroweak gauge
bosons and Goldstones do contribute. The contributing diagrams are
depicted in Fig.~2b. The contribution $\Sigma_{HH}^{tt}$
of the diagram of Fig.~2a has already been calculated in
the previous Section (\ref{Sigmatt1})--(\ref{Sigmatt2}).
In a completely analogous way, we obtain for the truncated
Green functions of Fig.~2b~\footnote{
Calculated in the Landau gauge, for reasons explained in the text following
eq.~(\ref{V1lgb}).}
\begin{equation}
\Sigma_{HH}^{ZG}(q^2) + \Sigma_{HH}^{WG}(q^2) =
 \frac{3}{16 \pi^2 \langle \varphi \rangle^2} q^2
\left( M^2_Z + 2 M^2_W \right) \left[ \ln \left( \frac{\Lambda^2}{m^2_t}
\right) + {\cal {O}}\left( \Lambda^0 \right) \right] \ .
\label{Sigmagb}
\end{equation}
Since $\delta Z^{\prime}_{\varphi} = - d \Sigma_{HH}/dq^2 |_{q^2=M^2_H}$
(cf.~eq.~(\ref{phiren})), this leads to the following modification
\begin{displaymath}
\Sigma_{HH} \ \mapsto \ \Sigma_{HH}^{tt} +
 \left( \Sigma_{HH}^{ZG}+ \Sigma_{HH}^{WG} \right) \qquad \Rightarrow
 \qquad
\delta Z_{\varphi} \mapsto \delta Z^{\prime}_{\varphi}  =
\kappa^{\prime} \left[ \ln \left( \frac{\Lambda^2}{m^2_t} \right)
+ {\cal {O}}\left( \Lambda^0 \right) \right] \ ,
\end{displaymath}
\begin{equation}
\mbox{where: } \quad \kappa^{\prime} = \kappa -
 \frac{3}{16 \pi^2} \frac{\left( M^2_Z+ 2 M^2_W \right)}{v^2}
= 1.37 \cdot 10^{-2} \ .
\label{Zprimephi}
\end{equation}
We note that we can obtain this result also very quickly by looking at
the 1-loop renormalization group equation (RGE) for the ``running'' VEV
$v(E) = \langle \varphi^{(E)} \rangle$. Such an RGE can be found, for 
example, in ref.~\cite{ackm}.

Furthermore, for the sake of consistency, we also include 
the 1-loop contributions of the electroweak gauge bosons to
the renormalization of the Yukawa coupling parameter $g_t$
(cf.~(\ref{Yukren})--(\ref{Zgl2})). They can most easily be obtained
by from the 1-loop RGE~\footnote{From
this RGE we see that, at 1-loop, there
are no scalar self-interaction contributions to the running of $g_t$.}
for the Yukawa coupling $g_t$
\begin{equation}
Z_{gt} \mapsto Z^{\prime}_{gt}  =  
+ 2 \delta Z_{gl} - \frac{3}{2} \delta Z_{\varphi} +
\frac{1}{48 \pi^2} \frac{\left( 17 M^2_Z + 10 M^2_W \right) }{v^2}
\left[ \ln \left( \frac{\Lambda^2}{m^2_t} \right) + {\cal {O}}\left(
\Lambda^0 \right) \right] \ ,
\label{Zprimegt}
\end{equation}
where $\delta Z_{gl}$ and $\delta Z_{\varphi}$ are given in (\ref{delZs}).
This could be obtained also in a more tedious way, by looking at the
1-loop contributions of the electroweak gauge bosons and Goldstones
to the top quark propagator and finding the corresponding change of the
pole mass (taking only the $\Lambda$-dependent part). This change would 
correspond to the gauge bosonic and Goldstone contribution in the combination
$(\delta Z^{\prime}_{\varphi} + \delta Z^{\prime}_{gt})/2$, since
$m_t \propto \langle \varphi \rangle g_t$. One may ask whether this
combination, or equivalently $\delta Z^{\prime}_{gt}$, is free of
$\Lambda^2$-terms (which are ignored by RGEs). The answer is yes, because
such terms come from tadpole diagrams and they cancel out due to the ``gap''
equation (\ref{1lgap}). The argument is closely analogous to that presented
in the previous Section after eq.~(\ref{mtren}).
All $\Lambda^2$-terms in the physical (pole) mass $m_t$ are already
contained in the bare mass factor $g_t(\Lambda) \langle
\varphi \rangle/ \sqrt{2}$, i.e., in the VEV $\langle \varphi \rangle$
of (\ref{1lgap}), and none are in the radiative correction factor
$1+(\delta Z^{\prime}_{\varphi} + \delta Z^{\prime}_{gt})/2$
(cf.~(\ref{mtren})).

In order to complete the Section, we must also find the physical (pole)
mass $M_H$ of the Higgs in terms of $\Lambda$ and the bare coupling
$\lambda(\Lambda)$. Now, $M_H$ must be
corrected by the 1-loop scalar and electroweak gauge bosonic effects.
The formula (\ref{MHpole1}) still applies, except that now we have to
calculate the second derivative of the effective potential modified
by the scalar and gauge bosonic contributions (\ref{V1lsc}) and
(\ref{V1lgb}), and the truncated Green function $\Sigma_{HH}^{tt}$
should be replaced by the full 1-loop truncated Green function
$\Sigma_{HH}$ with the gauge bosonic and scalar loop contributions
included
\begin{eqnarray}
M^2_H & = &
\frac{d^2}{d \varphi^2} \left( 
V_{\mbox{\footnotesize eff}}^{(0+1\ell t)} 
+ \delta V_{\mbox{\footnotesize eff}}^{(1\ell sc)} 
+ \delta V_{\mbox{\footnotesize eff}}^{(1\ell gb)} \right)
{\Big|}_
{\varphi= \langle \varphi \rangle} + \left[ \Sigma_{HH}\left(
q^2 =  M^2_H \right)
- \Sigma_{HH}\left( q^2 = 0 \right) \right] \ .
\label{MHpolem1}
\end{eqnarray}
It is straightforward to check that, again,
only the diagrams of Figs.~2a-2b contribute to the difference of
$\Sigma$'s in (\ref{MHpolem1}), i.e., $\Sigma_{HH}$ in (\ref{MHpolem1})
can be taken to be the sum of the epressions (\ref{Sigmatt2})
(with $m_t^{(0)}(\Lambda)$ there replaced by $m_t$) and
(\ref{Sigmagb}).
Furthermore, the second derivative in (\ref{MHpolem1}) can be calculated
again in the spirit of perturbation
\begin{displaymath}
 \frac{d^2}{d \varphi^2} \left(
V_{\mbox{\footnotesize eff}}^{(0+1\ell t)} 
+ \delta V_{\mbox{\footnotesize eff}}^{(1\ell sc)} 
+ \delta V_{\mbox{\footnotesize eff}}^{(1\ell gb)} \right)
{\Big|}_
{\varphi= \langle \varphi \rangle}  \approx
\end{displaymath}
\begin{equation}
\approx  \left[
\frac{d^2 V_{\mbox{\footnotesize eff}}^{(0+1\ell t)} }{d \varphi^2}
+ \delta \langle \varphi \rangle
\frac{d^3 V_{\mbox{\footnotesize eff}}^{(0+1\ell t)} }{d \varphi^3}
\right] {\Bigg|}_{\varphi =\langle \varphi \rangle_{1\ell t} } +
 \frac{d^2 \left(
\delta V_{\mbox{\footnotesize eff}}^{(1\ell sc)} \right) }{d \varphi^2}
{\Bigg|}_{\varphi =\langle \varphi \rangle_{1\ell t} }
+\frac{d^2 \left(
\delta V_{\mbox{\footnotesize eff}}^{(1\ell gb)} \right) }{d \varphi^2}
{\Bigg|}_{\varphi \approx v } \ .
 \label{MHq0}
 \end{equation}
Similarly as in (\ref{1lgap}), we used here only the first two terms in the
Taylor expansion of the second derivative $d^2 V_{\mbox{\footnotesize eff}}
^{(0+1\ell t)}/d\varphi^2$ around the value $\varphi = \langle \varphi
\rangle_{1\ell t}$. The second derivatives of the corrective
potentials $\delta V_{\mbox{\footnotesize eff}}^{(1\ell sc)}$ and
$\delta V_{\mbox{\footnotesize eff}}^{(1\ell gb)}$ are evaluated,
according to convenience, at one of the following similar values:
$\langle \varphi \rangle_{1\ell t}$
and $\langle \varphi_{\mbox{\footnotesize ren.}} \rangle = v$.
We replace $\delta \langle \varphi \rangle$ by
$\delta \langle \varphi \rangle^2/(2 \langle \varphi \rangle_{1\ell t})$,
and use relation (\ref{delVEV}) for $\delta \langle \varphi \rangle^2$.
We then directly calculate (\ref{MHq0}) in this way and with help of
the results obtained so far. Then we insert the obtained expression
into relation (\ref{MHpolem1}) and end up with the following
square of the physical mass of the Higgs
\begin{eqnarray}
M^2_H & = & \frac{\lambda(\Lambda) v^2}{3} \left[
1 - 2 \delta Z^{\prime}_{\varphi} + 
\delta Z_{\varphi} \frac{3 \cdot 4 m^2_t}{\lambda(\Lambda) v^2}
+ {\cal {O}}\left(\frac{\kappa g^2_t}{\lambda(\Lambda)} \right)
\right]
\nonumber\\
&&-\frac{3 v^2}{8 \pi^2} \left[ 
\frac{\lambda^2(\Lambda)}{9} \ln \frac{\Lambda^2}{m^2_t} +
\frac{M^4_Z}{v^4}\ln \frac{\Lambda^2}{M^2_Z} +
2 \frac{M^4_W}{v^4} \ln \frac{\Lambda^2}{M^2_W} \right] 
\left[1 + {\cal {O}}\left( \delta Z_{\varphi} \right) 
+ {\cal {O}} \left( 1/\ln \frac{\Lambda^2}{m^2_t} \right) \right]
\nonumber\\
&&+ {\cal {O}} \left( \frac{\kappa g_t^2 \Lambda^2}{16 \pi^2} \right) \ .
\label{MHphy}
\end{eqnarray}
The leading part of the $\Lambda^2$-dependence in $M^2_H$,
i.e., the terms 
${\cal {O}} \left( \Lambda^2 \lambda(\Lambda)/ (16 \pi^2) \right)$
and ${\cal {O}} \left( \Lambda^2 M^2_{g.b.}/ (16 \pi^2 v^2) \right)$,
turn out to have the coefficient equal to zero. The remaining 
$\Lambda^2$-terms are
suppressed by a factor which is at most of order $(\kappa g^2_t)/\pi^2$,
as indicated in the last line of (\ref{MHphy}). These terms may appear
and would have its origin partly in the (neglected) terms
$\Lambda^2 {\cal {O}}(\kappa g^2_t \varphi^2)/(32 \pi^2)$ of the
scalar-induced effective potential
$\delta V_{\mbox{\footnotesize eff}}^{(1 \ell sc)}$
of (\ref{V1lsc}). For $\Lambda \stackrel{<}{\approx} 1.5 \mbox{ TeV}$, 
the terms 
${\cal {O}} \left( \kappa g_t^2 \Lambda^2/\pi^2 \right)$
appear not to surpass the 
$\ln \Lambda$-terms of the second line in (\ref{MHphy}) and appear to
change the value of $M^2_H$ in such a case at most by ten percent.
A more detailed analysis should also include these
possible terms in $M^2_H$.
Here we will neglect them
and consider the formula (\ref{MHphy}) only as an approximation that
results in an estimated overall error of ten percent or less for $M_H$,
in the case of relatively low ultraviolet
cut-offs $\Lambda \leq 1.5 \mbox{ TeV}$ of Table 1.
At the end of this Section we will discuss other contributions
to the estimated overall error of formula (\ref{MHphy}).

At this point, we are able to calculate the ultraviolet upper bounds
$\Lambda^{\mbox{\scriptsize u.b.}}$ from
(\ref{Lub3}), and the (physical) Higgs masses $M_H$ from (\ref{MHphy})
-- both interrelated as functions of one single unknown variable, 
the bare scalar self-coupling parameter $\lambda(\Lambda)$.
The results are presented in the last two columns of Table 1. For comparison,
we also included the values when the gauge bosonic 1-loop effects were
neglected (fourth and fifth column). The values from Section 1, when
the 1-loop scalar self-interaction effects were neglected as well, are in
the second and third columns.

As emphasized at the beginning of the present Section, we confined
ourselves only to such values of $\lambda(\Lambda)$ for which the
calculated values have predictive power -- i.e., we confined
ourselves to the cases when the
1-loop scalar self-interaction effects on the effective potential
remain perturbative, changing (increasing) the square of the predicted
ultraviolet cut-off $\Lambda$ by roughly 50 percent or less. On the
other hand, it can be seen, by using (\ref{delVEV}), that the resulting
values for $\Lambda$ (Table 1, last column) change (decrease) the
square of the VEV typically by 50 percent or less:~\footnote{
In the extreme case of $\lambda(\Lambda) = 3.00$ in Table 1,
the square of the new VEV $\langle \varphi \rangle^2$ is decreased to
42 percent of the value of $\langle \varphi \rangle^2_{1\ell t}$, i.e.,
$\langle \varphi \rangle \approx 0.65 \langle \varphi \rangle_{1\ell t}$
in this case.}
$ |\delta \langle \varphi \rangle^2 |
= \langle \varphi \rangle^2_{1 \ell t} - \langle \varphi \rangle^2
\stackrel{<}{\approx} 0.5 \langle \varphi \rangle^2_{1 \ell t}$.
This means that the effects of
$\delta V^{(1\ell sc)}_{\mbox{\footnotesize eff}}$ and
$\delta V^{(1\ell gb)}_{\mbox{\footnotesize eff}}$ do not wash
out the VEV $\langle \varphi \rangle_{1 \ell t}$
to the value zero or almost zero. If they did,
they would make our assumption of the perturbative nature of these
effects non-viable in retrospect.

The philosophy followed in the present paper is essentially different
from that of the authors of~\cite{dp}-\cite{cp}. They assumed that
a new physics (at $E \stackrel{>}{\approx} \Lambda$) protects with a
symmetry the masses of the scalars and of the top quark from acquiring
$\Lambda^2$-dependent (and even $\ln \Lambda$-dependent,
cf.~\cite{lndp}-\cite{cp}) terms. In such a case, the tree level VEV
($\langle \varphi \rangle_0 = \sqrt{6 M^2(\Lambda^2)/\lambda(\Lambda)}$)
would still be drastically changed by the separate quantum contributions of
the top quark, the scalar and the electroweak gauge boson sectors. However,
these potentially huge contributions would largely cancel each other,
thus resulting in the possibility of having very high cut-off $\Lambda$, even
in the case of the tree-level SSB. If using the effective potential
of the present paper, the cancelation relations of~\cite{dp}-\cite{cp}
can be reproduced by requiring that the $\Lambda^2$ and $\ln \Lambda$-terms
appearing in the first derivative $dV_{\mbox{\footnotesize eff}}/
d \varphi^2|_{\varphi = \langle \varphi \rangle}$ of the entire
(0+1)-loop effective potential $(V_{\mbox{\footnotesize eff}}^{(0+1 \ell t)}
+ \delta V_{\mbox{\footnotesize eff}}^{(1 \ell sc)}
+ \delta V_{\mbox{\footnotesize eff}}^{(1 \ell gb)})$ are zero.
In the terminology of the present
paper, we would have in such a case essentially the cancelation of
the leading top quark and scalar self-interaction quantum effects:
$\langle \varphi \rangle^2_{1 \ell t} + \delta \langle \varphi \rangle^2
\approx \langle \varphi \rangle^2_0$. This would require the scalar
self-interactions to be stronger than those assumed in the present
paper, and therefore the Higgs mass would be heavier
($M_H \stackrel{>}{\approx} 300 \mbox{ GeV}$, cf.~\cite{cp}) than the
highest Higgs mass of Table 1 ($M_H \approx 235 \mbox{ GeV}$).

We stress that the results of Table 1 are valid only for the
case when the bare parameters $M^2(\Lambda)$ and $\lambda(\Lambda)$
of the starting tree level scalar potential (\ref{Vtree}) satisfy
either one of the two conditions (\ref{Lubco}) which can be rewritten
in the framework of this Section as:
\begin{equation}
\mbox{either: } \quad
0 \ \leq \ M^2(\Lambda) , \qquad
\mbox{or: } \quad
-M^2(\Lambda) = |M^2(\Lambda)| \ \ll \
\frac{\lambda(\Lambda)v^2}{6} \left[ 1 +
\frac{|\delta \langle \varphi \rangle^2 |}{v^2} \right] \
\left( = {\cal {O}}(M^2_H) \right) \ .
\label{Lubco2}
\end{equation}
Here, $\delta \langle \varphi \rangle^2 = - |\delta \langle \varphi \rangle^2|$
is given in (\ref{delVEV}) (note that $|\delta \langle \varphi \rangle^2|/v^2
\stackrel{<}{\approx} 1$ in the discussed cases).
Incidentally, this condition includes also
all cases of the (tree level) SSB in (\ref{Vtree}), i.e.,
$M^2(\Lambda) > 0$.

However, we point out that, similarly as argued at the end of the previous
Section, the results of the present Section basically survive even in the
case when
\begin{equation}
-M^2(\Lambda) = |M^2(\Lambda)| = {\cal {O}}(\lambda v^2)  \
\left( = {\cal {O}}(M^2_H) \right) \ .
\label{Lubco3}
\end{equation}
The values of $M^2_H$ in (\ref{MHphy}) do not have an explicit dependence
on $M^2(\Lambda)$ and therefore remain unchanged in such a case.
The values of $\Lambda^2$, however, are then modified:
they are then equal to the r.h.s.~of (\ref{Lub3}) amplified by the
following factor $k_{\ast}(\Lambda)$
\begin{equation}
k_{\ast}(\Lambda) = \left[ 1 + \frac{6 |M^2(\Lambda)|}{\lambda(\Lambda)v^2}
\left( 1 + \delta Z^{\prime}_{\varphi} \right) \right] \ .
\label{kast}
\end{equation}
This factor is then of order one. Hence, we have
$\Lambda = {\cal {O}}(1 \mbox{ TeV})$ also in the case of (\ref{Lubco3}),
although the numerical values of $\Lambda$'s of Table 1 are not valid then.

If, on the other hand, $M^2(\Lambda)$
is negative {\it and} its absolute value is larger than
$\lambda(\Lambda) v^2$ by at least an order of magnitude,
then we have a possibility to avoid the stringent upper bounds
of Table 1 for the ultraviolet cut-off $\Lambda$.
In such a case, as seen already from the ``gap'' equation (\ref{1ltgap1}),
we must have very large bare parameter $|M^2(\Lambda)| \approx
\kappa \Lambda^2 \sim 10^{-2} \Lambda^2 $, while the dimensionless
bare coupling $\lambda(\Lambda)$ remains small ($\lambda(\Lambda)
\sim M^2_H/v^2$). A special case of this,
namely the case when the scalar doublet is not self-interacting at the
cut-off scale $\Lambda$ and has a finite bare mass there
($\lambda(\Lambda) \approx 0$ and $\mu^2(\Lambda)=-M^2(\Lambda) > 0$) was
discussed in ref.~\cite{fghw}. The authors of ref.~\cite{fghw} obtained
in such a case a wide range of possibilities for the values of $\Lambda$
($1 \mbox{ TeV} \stackrel{<}{\sim} \Lambda \stackrel{<}{\sim} E_{\mbox
{\scriptsize GUT}}$). They obtained $\Lambda \approx 1 \mbox{ TeV}$
(and the physical mass $M_H \approx 80 \mbox{ GeV}$) only if, in addition, they
demanded that the fourth derivative of $V_{\mbox{\footnotesize}}
^{(0+1 \ell t)}$ be zero at the VEV. The latter requirement gives in the
case of non-zero (positive) small $\lambda(\Lambda)$ a higher Higgs mass
(cf.~(\ref{MHpole2rf}) and (\ref{MHpole3})), but an even lower 
cut-off because
\begin{equation}
\frac{d^4 V_{\mbox{\footnotesize eff}}^{(0+1 \ell t)} }{d \varphi^4}
{\Big |}_{\varphi = \langle \varphi \rangle_{1 \ell t}} =
\lambda(\Lambda) + 6 \kappa g^2_t \left[ 
\ln\frac{\Lambda^2}{m^2_t}
-\frac{11}{3} + {\cal {O}}\left(\frac{m^2_t}{\Lambda^2} \right)
\right] \ .
\label{fourth}
\end{equation}

At the end, we discuss the errors involved in the upper bounds
$\Lambda^{\mbox{\scriptsize u.b.}}$ of
Table 1 (seventh column). The errors appear mostly due to
our neglecting those terms on the r.h.s.~of (\ref{Lub3}) which
we denoted there with ${\cal {O}}(\cdots)$, and because we neglected
possible terms of ${\cal {O}}(\Lambda^0)$ in the $\ln \Lambda$-dominated
expressions for $\delta Z_{gl}$ and $\delta Z_{\varphi}$. The
errors for the upper bound of $\Lambda^2$
resulting from the latter approximation are estimated to
be at most 9-10 percent (we take: ${\cal {O}}(\Lambda^0) = 1$ for
estimate). The bulk of this error (up to 7 percent) comes from the
uncertain ${\cal {O}}(\Lambda^0)$-term of the gluonic (QCD)
renormalization effect $\delta Z_{gl}$ (\ref{delZs}). The terms
denoted as ${\cal {O}}(\kappa g^2_t/\lambda(\Lambda))$ in
(\ref{Lub3}) are magnified by a factor of order 10,
as explicit
preliminary calculations of these terms indicate.
They then result in an error of up to 5 percent for the upper bound of
$\Lambda^2$. The terms ${\cal {O}}(\theta(\Lambda))$
in (\ref{Lub3}) also result in an error of up to 5 percent. The errors
from the uncertainties of the last term in the curly brackets
of (\ref{Lub3}) (i.e., of the gauge bosonic contributions) are
quite negligible. Finally, the perturbative approximation of
(\ref{1lgap}), where we included only the leading term in the
expansion of $d(\delta V^{(1\ell sc)}_{\mbox{\scriptsize eff}})/
\delta \varphi^2$ around the value of $\varphi^2 = \langle \varphi
\rangle^2_{1\ell t}$, would also amount to a certain error in
$\delta \langle \varphi \rangle^2$, and therefore in the upper bound
for $\Lambda^2$. Explicit preliminary calculations, where
we include also the next term in the Taylor expansion, indicate that
the error committed in $\delta \langle \varphi \rangle^2$ is about
15-20 percent for the values of $\Lambda$ of Table 1. This in turn
would result in an error for the upper bound on $\Lambda^2$
of 8-12 percent.

Therefore, regarding all the sources of the discussed errors
as being independent, we end up with a rough estimate of 10-18 percent
for the possible error in the upper bound of $\Lambda^2$.
This implies an estimated error of 5-10 percent for the values of the
upper bound of $\Lambda$ in the last column of Table 1.

For large values of $\lambda(\Lambda)$ ($\lambda(\Lambda) > 2$),
we have, in addition, uncertainties arising from the ``higher order''
contributions $\propto \lambda^3(\Lambda)$, i.e., contributions that
include also the 2-loop
effects of the scalar self-interactions. They result in relative
corrections to the upper bound of $\Lambda^2$ which we conjecture to be
roughly of the order of $[ \lambda(\Lambda)/(32 \pi^2 \kappa)]^2$, i.e.,
of the order of the next (deleted) term of the geometric series in
(\ref{Lub3}). Hence, these corrections are appreciable in the cases
of the upper half of the values of
$\lambda(\Lambda)$ taken in Table 1. For example, for $\lambda(\Lambda) =$ 
2.25, 3.00, these corrections to the upper bound of $\Lambda^2$
would be roughly 12 and 22 percent, respectively; such effects
would then increase the previously estimated uncertainty of 5-10 percent
for $\Lambda^{\mbox {\scriptsize u.b.}}$ to 8-12 and 12-15 percent,
respectively. The analogous uncertainties of the gauge bosonic effects
are negligible.

We wish to point out, however, that the estimated errors for the
cut-off $\Lambda$ should be looked upon with some reservation, connected
with the inherently approximate nature of the meaning of $\Lambda$
as the scale beyond which new physics emerges. Realistically, there
is no such thing as a definite, abrupt cut-off. The transition to
the new physics is gradual as we increase the energy of probes.

On the other hand, the uncertainties in the expression (\ref{MHphy})
for $M^2_H$ coming from the first term (the square brackets) can be
estimated to contribute up to 8 percent, and the uncertainties of the terms
which are proportional to $\ln \Lambda$ in (\ref{MHphy})
contribute up to 5 percent. As argued just after eq.~(\ref{MHphy}),
the terms of ${\cal {O}}(\kappa g^2_t \Lambda^2/\pi^2)$ would change
$M^2_H$ by up to 10 percent (for $\Lambda < 1.5 \mbox{ TeV}$).
Altogether, this would result in an estimated uncertainty of 
10-15 percent for $M^2_H$, and correspondingly 5-8 percent for
$M_H$ of Table 1 (sixth column). The uncertainty due to
our ignoring the 2-loop contributions of scalars and gauge bosons
to $M_H$ appear to be substantially lower, even for the
case of large $\lambda (\Lambda) \approx 3$.

\section{Conclusions}
We found out that in the minimal Standard Model,
for a large subsector (\ref{Lubco2})-(\ref{Lubco3}) of the possible
values of the bare parameters $M^2(\Lambda)$ and $\lambda(\Lambda)$
in the tree level potential (\ref{Vtree}), 
the ultraviolet cut-off $\Lambda$ of the theory should not be
larger than ${\cal {O}}(1 \mbox{ TeV})$, as long as we
demand that the scalar self-interactions not be too strong,
i.e., that they behave perturbatively. By the latter we mean that
the 1-loop contributions of the scalar self-interaction to the (derivative
of the) effective potential are taken to be distinctly smaller than
those of the heavy quark Yukawa interaction. Furthermore, it turns out
that the corresponding contributions of the electroweak gauge bosons are
substantially smaller than those of the heavy top quark.
The resulting Higgs masses are in the range 150-250 GeV.

The heavy top quark
and the corresponding $\Lambda^2$-terms in the top-induced
effective potential play
a crucial role leading to the conclusions of the
present paper. These effects of the heavy top quark sector on the
scalar one are inherently non-perturbative.
The case of the tree-level SSB ($M^2(\Lambda) > 0$) is just
one part of the subsector (\ref{Lubco2}) leading to 
$\Lambda \sim 1\mbox{ TeV}$. The other case 
of (\ref{Lubco2})-(\ref{Lubco3})
where the conclusions of the paper apply is: no
tree-level SSB, with massless or not very heavy scalar doublet at the tree
level whose bare mass is
$\mu(\Lambda) = \sqrt{-M^2(\Lambda)} \leq
{\cal {O}}(v \sqrt{\lambda(\Lambda)}) \ (= {\cal {O}}(M_H) )$.
If the bare parameters $M^2(\Lambda)$ and $\lambda(\Lambda)$
do not fulfill any of the conditions (\ref{Lubco2})-(\ref{Lubco3}), i.e.,
if $\mu(\Lambda) = \sqrt{-M^2(\Lambda)}$ is by at least an order of
magnitude larger than $v \sqrt{\lambda(\Lambda)} = {\cal {O}}(M_H)$, then
the cut-off $\Lambda$ can become higher than ${\cal {O}}(1 \mbox{ TeV})$,
roughly of the order of
$\mu(\Lambda)/\sqrt{\kappa} \sim 10 \mu(\Lambda)$. The implications for the
values of the cut-off $\Lambda$ remain unclear if the scalar sector is
strongly interacting, i.e., if the Higgs has a substantially 
larger mass than the values $M_H$ listed in Table 1.

\section*{Acknowledgment}
This work was supported in part by the Deutsche Forschungsgemeinschaft
(DFG). The author wishes to thank A.~Kehagias for pointing out the
importance of heavy top quark for the scalar sector.

\section{Note added}
After finishing the present work, a related work of T.~Hambye
came to our attention~\cite{hamb}. He discusses the case of the
zero bare mass parameter $\mu^2(\Lambda)=0$. We note that, in the
present paper, the cut-off $\Lambda$ acquires the values
$\Lambda^{\mbox{\footnotesize u.b.}}$ of Table 1 precisely in
that case. However, Hambye does not
assume that the top quark loops necessarily dominate over the
1-loop scalar self-interaction effects in the effective potential.
Thus, he allows for the values of the bare coupling $\lambda(\Lambda)$ 
a larger region which would be restricted only by the requirement that 
the scalar self-interaction be weak enough to be treated 
perturbatively. 
In the region where $\lambda(\Lambda)
\leq 3$, his values for $\Lambda$ and $M_H$ agree roughly
with the corresponding results of the present paper, i.e.,
with $\Lambda^{\mbox{\footnotesize u.b.}}(1lt+sc+gb)$ and
$M_H(1lt+sc+gb)$ of Table 1, respectively.
The small differences in numbers (for $\lambda(\Lambda) \leq 3$)
arise largely from the fact that we took in account, in addition,
the effects of the running of the various 
discussed parameters ($g_t$, $m_t$, $\varphi$,
$M_H$) from the cut-off $E=\Lambda$ down to the electroweak energies.

\newpage

\vspace{4cm}

\begin{table}[h]
\vspace{0.3cm}

\begin{center}
Table 1 \\
\vspace{0.5cm}
\begin{tabular}{|c|c|c|c|c|c|c|} \hline
$\lambda(\Lambda)$ & $M_H^{(1\ell t)}$ & 
$\Lambda^{\mbox{\scriptsize u.b.}} (1\ell t)$ &
$M_H (1\ell t+sc)$ & 
$\Lambda^{\mbox{\scriptsize u.b.}} (1\ell t+sc)$ &
$M_H (1\ell t+sc+gb)$ & 
$\Lambda^{\mbox{\scriptsize u.b.}} (1\ell t+sc+gb)$ \\ 
 & [GeV] & [TeV] & [GeV] & [TeV] & [GeV] & [TeV] \\ \hline
3.00 & 247 & 0.94 & 230 & 1.16 (51 {\%})   & 234 & 1.24 \\
2.75 & 238 & 0.90 & 223 & 1.09 (46 {\%})   & 228 & 1.18 \\
2.50 & 229 & 0.87 & 217 & 1.03 (42 {\%})   & 221 & 1.11 \\
2.25 & 219 & 0.83 & 209 & 0.97 (38 {\%})   & 213 & 1.04 \\
2.00 & 208 & 0.78 & 201 & 0.90 (34 {\%})   & 204 & 0.98 \\
1.75 & 197 & 0.74 & 191 & 0.84 (29 {\%})   & 194 & 0.90 \\
1.50 & 185 & 0.69 & 181 & 0.77 (25 {\%})   & 183 & 0.83 \\
1.25 & 171 & 0.64 & 169 & 0.70 (21 {\%})   & 171 & 0.75 \\
1.00 & 156 & 0.58 & 155 & 0.62 (17 {\%})   & 157 & 0.67 \\ \hline
\end{tabular}
\end{center}
\end{table}

\clearpage

\oddsidemargin-0.5cm
\evensidemargin-0.5cm

\section{Table and figure captions}

\noindent {\bf Table 1}: The upper bounds for the ultraviolet
cut-off $\Lambda$ as a function of the physical Higgs mass, or
alternatively, of the bare scalar self-interaction parameter 
$\lambda(\Lambda)$. The second and third column refer 
to the values when only the leading heavy top quark 
quantum contributions (non-perturbative) are included in 
the effective potential. The fourth and fifth columns are for 
the case when the leading scalar self-interaction quantum
contributions (perturbative) are included; in the fifth column, we
included in parentheses the percentages by which the upper bound
for the square of the ultraviolet cut-off ($\Lambda^2$) changed with 
respect to the values of the third column. 
The sixth and seventh column are for the values
when, in addition, also the leading electroweak gauge boson quantum
contributions (perturbative) are included. We took $m_t^{\mbox
{\scriptsize phy}} = 180 {\mbox { GeV}}$, and always included also
the leading logarithmic QCD correction to the running of the Yukawa
coupling $g_t$ between $m_t$ and $\Lambda$.

\vspace{1cm}

\noindent {\bf Figures 1a-c}: The 1-PI diagrams whose truncated
Green functions lead to
the 1-loop heavy quark contributions to the effective potential
(calculated with the notation of unbroken fields).

\vspace{1cm}

\noindent {\bf Figures 2a-b}: The diagrams whose truncated Green
functions $\Sigma_{HH}^{tt}(q^2)$, $\Sigma_{HH}^{ZG}(q^2)$ and
$\Sigma_{HH}^{WG}(q^2)$ contribute in the Landau gauge
non-zero $\ln \Lambda$-terms to $[d \Sigma/dq^2|_{q^2=M^2_h}]$ and to
$[\Sigma_{HH}(M^2_H) - \Sigma_{HH}(0)]$, and hence contribute
to the renormalization of the field and the mass of the Higgs.
The Goldstones are denoted as $G^{(0)}$ and $G^{(\pm)}$.

\vspace{1cm}

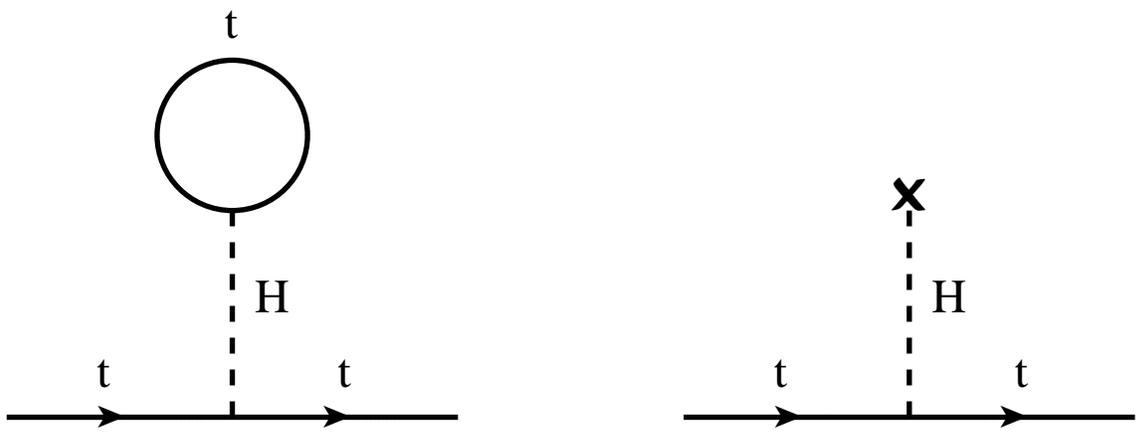
\noindent {\bf Figure 3}: The tadpole contributions to the pole
mass of the top quark, when 1-loop scalar self-interaction effects
and gauge bosons are ignored. The mass in the tree-level propagators
of the top quark is the bare mass $m_t^{(0)}(\Lambda)$; H is the
Higgs field ($H=\varphi - \langle \varphi \rangle_{1 \ell t}$);
the second diagram is the contribution of the linear $H$-term.


\vspace{4cm}

\begin{thebibliography}{99}

\bibitem{sopczak}
A.~Sopczak, CERN preprint CERN-PPE/95-46, hep-ph/9504300; 
M.~Consoli and Z.~Hioki,
{\it Mod.~Phys.~Lett.}~{\bf A10}, 2245 (1995);
J.~Ellis, G.L.~Fogli and E.~Lisi, preprint CERN-TH/95-202,
hep-ph/9507424; P.H.~Chankowski and S.~Pokorski,
Max-Planck-Inst.~preprint MPI-Ph/95-39, hep-ph/9505304 and update
hep-ph/9509207.

\bibitem{CDF}
F.~Abe et al., CDF Collaboration, {\it Phys.~Rev.~Lett.}~{\bf 72}, 225 
(1994); {\it ibid.}~{\bf 75}, 3997 (1995).


\bibitem{bhl}
V.~A.~Miranski, M.~Tanabashi and K.~Yamawaki, 
{\it Mod.~Phys.~Lett.}~{\bf A4}, 1043 (1989);
{\it Phys. Lett.}~{\bf B221}, 177 (1989); W.~A.~Bardeen, 
C.~T.~Hill and M.~Lindner, {\it Phys.~Rev.}~{\bf D41}, 1647 (1990).

\bibitem{fghw}
J.~P.~Fatelo, J.-M.~G\'erard, T.~Hambye and J.~Weyers,
{\it Phys.~Rev.~Lett.}~{\bf 74}, 492 (1995).

\bibitem{gceap}
G.~Cveti\v c and E.A.~Paschos, {\it Nucl.~Phys.}~{\bf B395}, 581 (1993).

\bibitem{huang}
K.~Huang: Quarks, Leptons and Gauge Fields
(World Scientific, Singapore, 1982);
D.~Bailin and A.~Love: Introduction to Gauge Field Theory
(Adam Hilger, Bristol and Boston, 1986).

\bibitem{ackm}
H.~Arason, D.J.~Casta\~no, B.~Kesthelyi, S.~Mikaelian, E.~J.~Piard,
P.~Ramond and B.~D.~Wright, {\it Phys.~Rev.}~{\bf D46}, 3945 (1992),
Appendix A.

\bibitem{dp}
R.~Decker and J.~Pestieau, Louvain University preprint UCL-IPT-79-19
(1979); DESY Workshop Oct.~22-24, 1979;
{\it Mod.~Phys.~Lett.}~{\bf A4}, 2733 (1989);
{\it ibid.}~{\bf A7}, 3773 (1992);
M.~Veltman, {\it Acta Phys.~Pol.}~{\bf B12}, 437 (1981);
M.~S.~Al-sarhi, I.~Jack and D.~R.~T.~Jones, {\it Z.~Phys.}~{\bf C55}, 283 
(1992);
and references therein.

\bibitem{lndp}
J.~Pestieau and P.~Roy, {\it Phys.~Rev.~Lett.}~{\bf 23}, 349 (1969);
R.~Decker and J.~Pestieau, {it Lett.~Nuovo Cim.}~{\bf 29}, 560 (1980);
P.~Osland and T.~T.~Wu, {it Z.~Phys.}~{\bf C55}, 569, 585, 593 (1992);
and references therein.

\bibitem{cp}
G.~L\'opez Castro and J.~Pestieau, {\it Mod.~Phys.~Lett.}~{\bf A10}, 1155 
(1995).

\bibitem{hamb}
T.~Hambye, ``Symmetry breaking induced by top quark loops from a model
without scalar mass'', Louvain preprint UCL-IPT-95-16, hep-ph/9510266,
to appear in {\it Phys.~Lett.}~{\bf B} (update 17 Apr.~96: 
T.~Hambye, {\it Phys.~Lett.}~{\bf B371}, 87 (1996)).



\end{thebibliography}
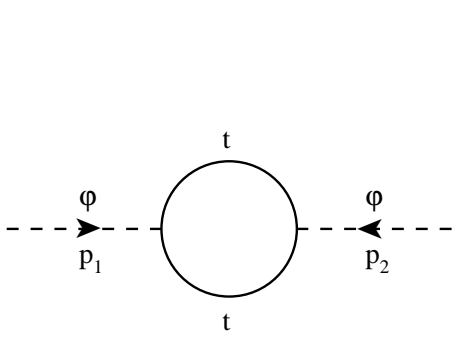
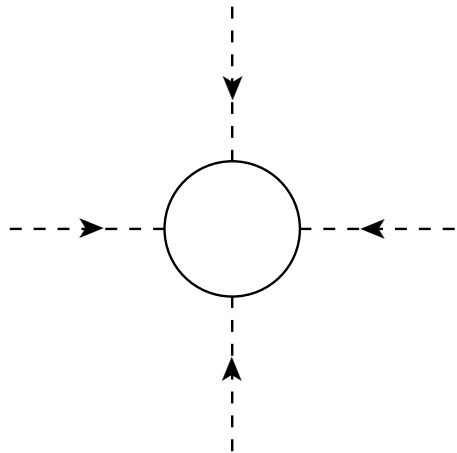
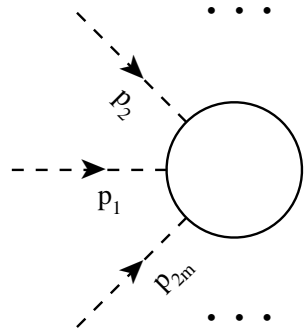
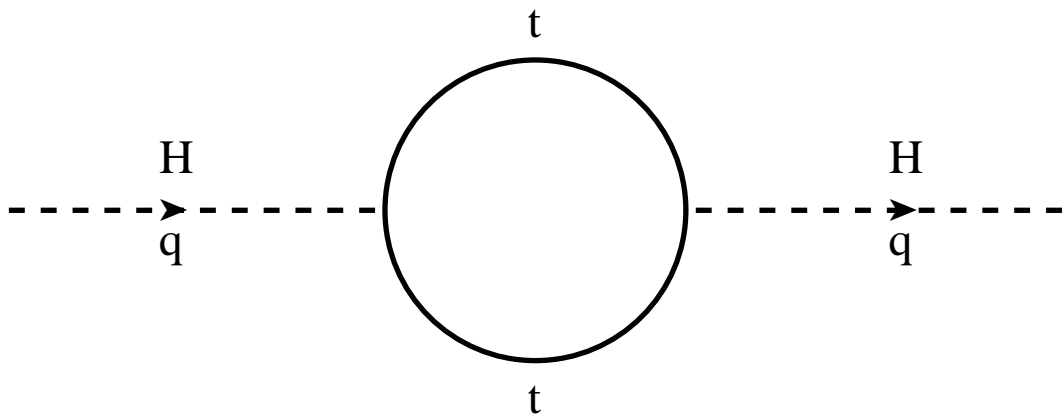
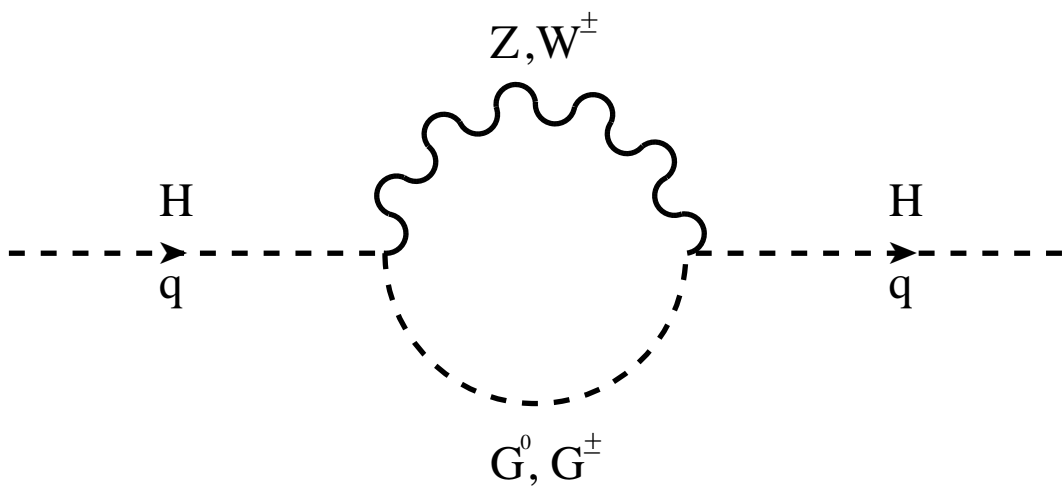
\end{document}